\documentclass[a4paper,5p, sort&compress, preprint]{elsarticle}

\usepackage[utf8]{inputenc}
\usepackage{graphicx}
\usepackage{float}
\usepackage{hyperref}
\usepackage{isotope} 
\usepackage{amsmath}
\usepackage{upgreek}
\usepackage{sidecap}
\usepackage{booktabs}
\usepackage{multirow}

\usepackage{siunitx}
\journal{Nucl.Instrum.Meth. A}

\begin{document}

\begin{frontmatter}

\title{Characterization of a large CdZnTe coplanar quad-grid semiconductor detector \\ \vspace{1em}
	\normalsize{The COBRA collaboration}\vspace{-1.5em}}

\author[Hamburg]{J.~Ebert} 
\author[Dresden]{D.~Gehre} 
\author[Dortmund]{C.~G\"o\ss{}ling} 
\author[Hamburg]{C.~Hagner} 
\author[Hamburg]{N.~Heidrich} 
\author[Dortmund]{R.~Klingenberg} 
\author[Dortmund]{K.~Kr\"oninger} 
\author[Dortmund]{C.~Nitsch} 
\author[Hamburg]{C.~Oldorf} 
\author[Dortmund]{T.~Quante} 
\author[Dortmund]{S.~Rajek} 
\author[Hamburg]{H.~Rebber} 
\author[Dresden]{K.~Rohatsch} 
\author[Dortmund]{J.~Tebr\"ugge} 
\author[Dortmund]{R.~Temminghoff} 
\author[Dortmund]{R.~Theinert\corref{cor1}} \ead{robert.theinert@tu-dortmund.de}
\author[Hamburg]{J.~Timm} 
\author[Hamburg]{B.~Wonsak} 
\author[Dresden]{S.~Zatschler} 
\author[Dresden]{K.~Zuber} 

\address[Hamburg]{Universit\"at Hamburg, Institut f\"ur Experimentalphysik Luruper Chaussee 149, 22761~Hamburg}
\address[Dresden]{Technische Universit\"at Dresden, Institut f\"ur Kern- und Teilchenphysik Zellescher Weg 19, 01069 Dresden}
\address[Dortmund]{Technische Universit\"at Dortmund, Lehrstuhl f\"ur Experimentelle Physik IV Otto-Hahn-Str.~4, 44221 Dortmund}

\cortext[cor1]{Corresponding author. Tel 49 231 755 3690; fax 49 231 755 3688.}

\pdfinfo{%
  /Title    (Characterization of a large CdZnTe coplanar quad-grid semiconductor detector)
  /Author   (R. Theinert)
  /Creator  ()
  /Producer ()
  /Subject  ()
  /Keywords (CZT \sep CdZnTe \sep semiconductor detector \sep coplanar \sep background reduction \sep double beta-decay \sep $\gamma$-ray spectroscopy)
}

\begin{abstract}
The COBRA collaboration aims to search for neutrinoless double beta-decay of $^{116}$Cd. A demonstrator setup with 64 CdZnTe semiconductor detectors, each with a volume of 1\,cm$^3$, is currently being operated at the LNGS underground laboratory in Italy.
This paper reports on the characterization of a large 2\:$\times$\:2\:$\times$\:1.5\,cm$^3$ CdZnTe detector with a new coplanar-grid design for applications in $\gamma$-ray spectroscopy and low-background operation. 
Several studies of electric properties as well as of the spectrometric performance, like energy response and resolution, are conducted. Furthermore, measurements including investigating the operational stability and a possibility to identify multiple-scattered photons are presented.
\end{abstract}


\end{frontmatter} 


\section{Introduction} \label{sec:introduction}
The COBRA collaboration operates a demonstrator setup at the Laboratori Nazionali del Gran Sasso (LNGS) in Italy to prepare a large-scale experiment for the search for neutrinoless double beta-decays of several isotopes using CdZnTe semiconductor detectors \cite{cobra01,cobra_demonstrator15}. 
A description of the current status of the experiment can be found in Ref. \cite{cobra_status13}.
Currently, 64 detectors with a volume of \SI[product-units = power]{1x1x1}{\centi\meter} and a mass of approximately \SI{5.9}{\gram} each are operated. 

Double beta-decays are associated with half-lives of more than \SI{e25}{y} \cite{rodejohann_0nbb11}. 
The sensitivity of the experiment on those half-lives is, among other aspects, driven by a high detection efficiency and an ultra low-background setup.
A promising improvement of the sensitivity is expected by using larger detectors.

Detectors with a larger volume have a higher full energy peak efficiency compared to the smaller \SI{1}{\centi\meter\cubed} detectors or thin pixel read out detectors. Thus, the probability for electrons, coming from a neutrinoless double beta-decay, to escape the detector is lower and the detection efficiency higher. 
The detection efficiency contributes linearly to the sensitivity and thus, the latter can be highly improved by using \SI{6}{\centi\meter\cubed} detectors.

Due to the larger mass of each detector, a smaller number of detectors is needed to reach a good sensitivity in a reasonable measuring time span. Less detectors lead to a lower background induced by cables, support material and readout electronics. Furthermore, the surface contribution induced by $\alpha$-contamination in the background is reduced as well due to a smaller surface-to-volume ratio.
Thus, the background can be more efficiently suppressed, which is, until now, a limiting factor of the sensitivity.

In the past, several groups investigated CdZnTe copla\-nar-grid (CPG) detectors with a volume larger than the commonly used \SI{1}{\centi\meter\cubed}. 
These include the investigation of an \SI{11.98}{\centi\meter\cubed} CdZnTe detector \cite{he_multiPair05}. These detector showed a degraded spectroscopic performance due to frequent electron trapping. Also, a \SI{5.45}{\centi\meter\cubed} CdZnTe detector and a \isotope{LaBr_3(Ce)} detector were compared in Ref. \cite{syntfeld_largeDet06}. It was pointed out that these large CdZnTe detector had a poor energy resolution and that more investigations were required until such detectors could be used for spectroscopy. The homogeneity of a \SI{6}{\centi\meter\cubed} detector was studied in Ref. \cite{bolotnikov_largeDet11}. It was found that areas with \isotope{Te}-inclusions cause a poor energy resolution. 

The development of CdZnTe semiconductor detectors is proceeding continuously and larger detectors with a better energy resolution are becoming commercially available. However, it is still difficult to grow large homogeneous detector-grade crystals which are significantly larger than \SI{1}{\centi\meter\cubed} and which meet requirements for usage in low-back\-ground operation.

This paper reports a characterization of a \SI[product-units = power]{2x2x1.5}{\centi\meter} CdZnTe semiconductor detector with four separated CPGs on the anode side manufactured by Redlen, Saanichton, Canada. 
All four grids are read out simultaneously, however, each grid is calibrated separately. 
Furthermore, the granularity offers the possibility to identify multiple-scat\-tered photons, which are an important source of background for low-background experiments.

This paper is structured as follows.
Section \ref{sec:concept} gives a short overview on CdZnTe semiconductor detectors, the CPG layout in general and the concept of coplanar quad-grid (CPqG) detectors as well as the readout electronics used. The dependency of the current and the capacitance on the applied voltage is studied in Section \ref{sec:det_characteristics}. In Section \ref{sec:performance}, the reconstruction of the energy, the linearity of the energy calibration, the energy resolution as well as further properties of the detector are presented. Section \ref{sec:simulations} describes Monte Carlo simulations of the setup. 
In Section \ref{sec:stability}, the detector is tested for its operational stability.  
Section \ref{sec:grid_correlation} shows the results for the detector studied to identify multiple-scattered photons.
Section \ref{sec:conclusion_outlook} concludes the paper.

\section{Detector concept and experimental setup} \label{sec:concept}
CdZnTe semiconductors have a high resistivity of approximately \SI{3e10}{\ohm\centi\meter} as well as a large band gap of about \SI{1.6}{e\volt} and can thus be operated at room temperature.
The products of the mobility and lifetime for electrons and holes are in a range of \SIrange[range-phrase = --,range-units = brackets,fixed-exponent = -3,scientific-notation = fixed]{4e-4}{11e-3}{\centi\meter\squared\per\volt} and \SIrange[range-phrase = --,range-units = brackets,fixed-exponent = -5,scientific-notation = fixed]{3e-6}{9e-5}{\centi\meter\squared\per\volt}, respectively. For summary of measurements see Ref. \cite{schlesinger_review01} and references therein.
Due to thus large difference, holes are trapped within a shorter distance than electrons. As the signal is induced by moving charge carriers trapped holes do not contribute to the signal. As a consequence, the energy measured with planar detectors depends on the interaction depth.
 
\begin{figure}[h!]
    \centering
    \includegraphics[width=0.62\columnwidth, angle=0]{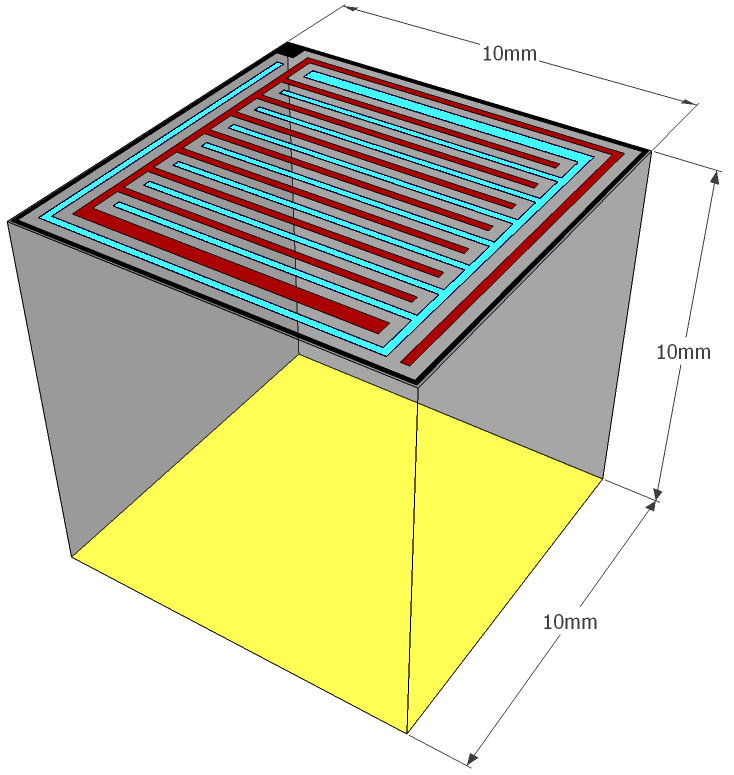}
    \caption[]{Scheme of a small \SI{1}{\centi\meter\cubed} coplanar-grid detector. Note the comb-shaped interleaved anode electrodes in dark red and light blue, and the planar cathode in light yellow on the opposite side.}
    \label{fig:CPG_small_det}
\end{figure}

Several techniques have been developed in the past to compensate for this \cite{he_ramo01}. For the present paper, a CPG design for the detector is used, which was first investigated in Ref. \cite{luke94}. The design shown in Figure \ref{fig:CPG_small_det} is similar to that of Frisch grids, employed in gaseous detectors.

The CPG detector features a cathode at negative high voltage. The anode side consists of two planar comb-shaped interleaved electrodes, out of which one, the \textit{collecting anode} (CA), is set to ground potential, while the other, the \textit{non-collecting anode} (NCA), is set to a slightly negative voltage. 
The voltage difference between both anodes is referred to as grid bias ($U_{\text{GB}}$), the voltage between the anodes and the cathode is referred to as bias voltage ($U_{\text{BV}}$). 

\begin{figure}[h!]
    \centering
    \includegraphics[width=0.62\columnwidth, angle=0]{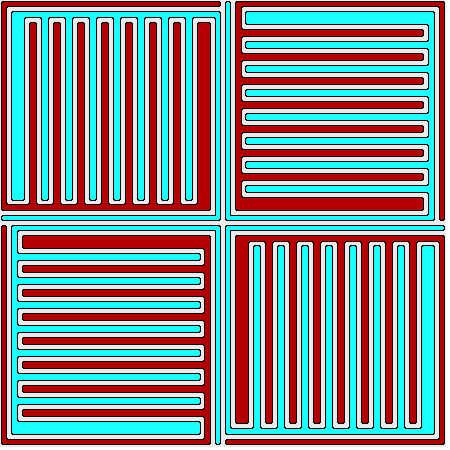}
    \caption[]{Design of the anode structure with its four sectors. The CAs are indicated in dark red and the NCAs in light blue.}
    \label{fig:CPG_quad_concept}
\end{figure}

The detector investigated for this paper has an anode grid structure composed of four CPGs, referred to as sectors, which are rotated against each other by \SI{90}{\degree} (see Fig. \ref{fig:CPG_quad_concept}). In comparison to a layout with four parallel grids (presented in Ref. \cite{he_multiPair05,ma_CPGarray2014}), the rotated design allows to generate the same electric field conditions on all lateral surface sides by applying the same electrical potential on all outer or inner electrodes. Such a grid configuration results in a homogeneous electric field in large parts of the detector between the cathode and the anodes.

For the detector under study, the outer four grids are operated as CAs, while the inner four grids are operated as NCAs. This offers the possibility of separating the four single sectors from each other by small non-collecting areas (virtual steering grid).

\begin{figure}[h!]
    \centering
    \includegraphics[width=0.62\columnwidth, angle=0]{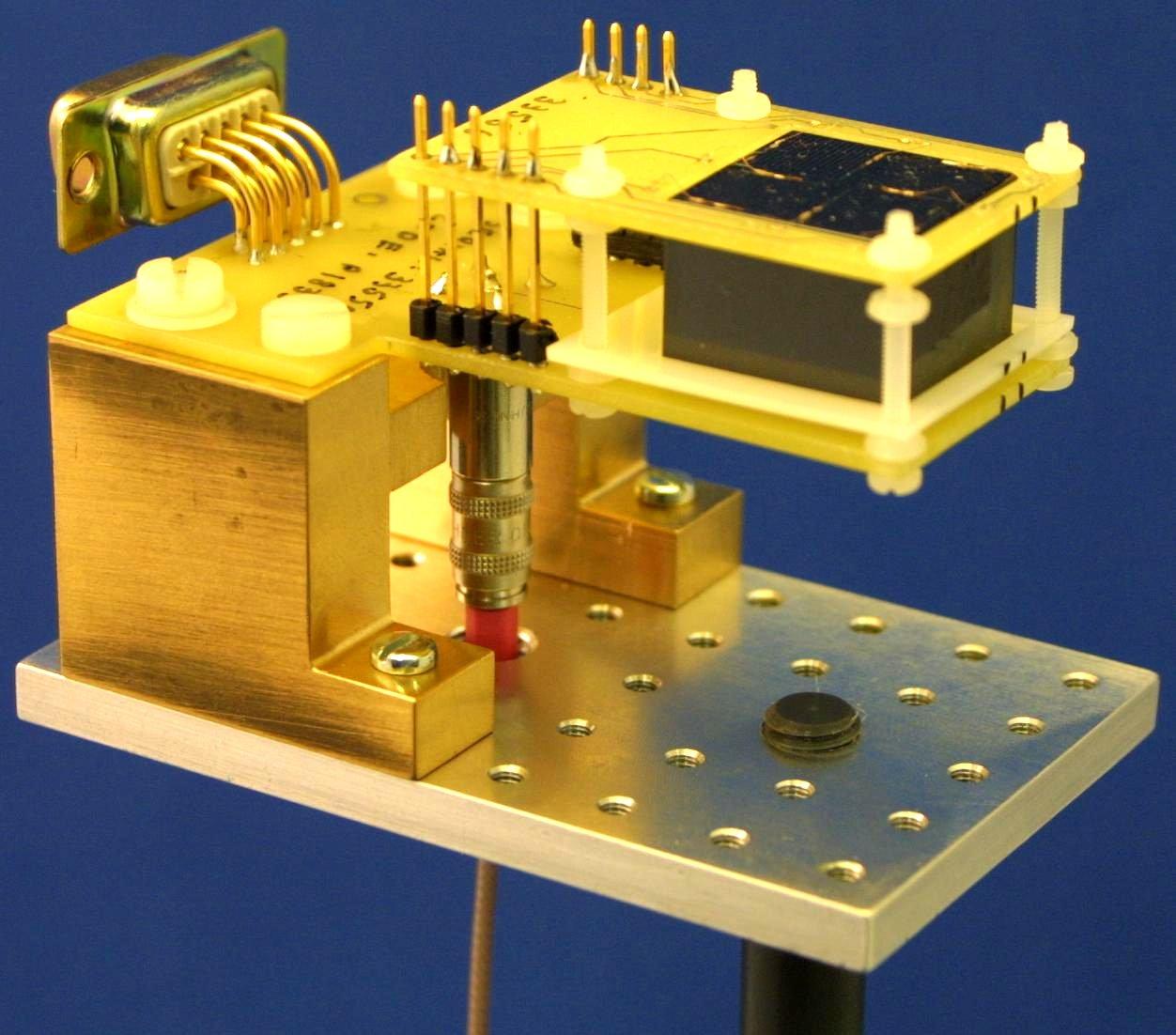}
    \caption[]{The contacted CPqG detector (upper right) used for the studies presented in this paper with its mounting PCB.}
    \label{fig:CPG_quad_setup}
\end{figure}

Figure \ref{fig:CPG_quad_setup} shows the CdZnTe crystal and the mounting system. The eight anodes are connected to contacting pads on the printed circuit board (PCB) via thin wires and a silver based conductive lacquer. The cathode is placed on the high voltage pad of the board and held in place by its net weight. A polyoxymethylene frame ensures stable positioning.

The raw signals from the detector are amplified by charge sensitive preamplifiers, which convert the charge signals into voltage signals. Furthermore, the single ended signals (SE) are converted to differential (diff) signals to have a robust transmission over the distance of about two meters from the preamplifier to the linear amplifier in the main electronic rack. 
The linear amplifier converts the signal to the operational range of the following flash analog to digital converter (FADC), which records the complete pulse shape with a sampling frequency of \SI{100}{\mega\hertz}. A computer monitors the measurement and stores the acquired data for further analysis. Details of the system can be found in Ref. \cite{diss_schulz,cobra_demonstrator15}.

\begin{figure}[h!]
    \centering
    \includegraphics[width=0.99\columnwidth, angle=0]{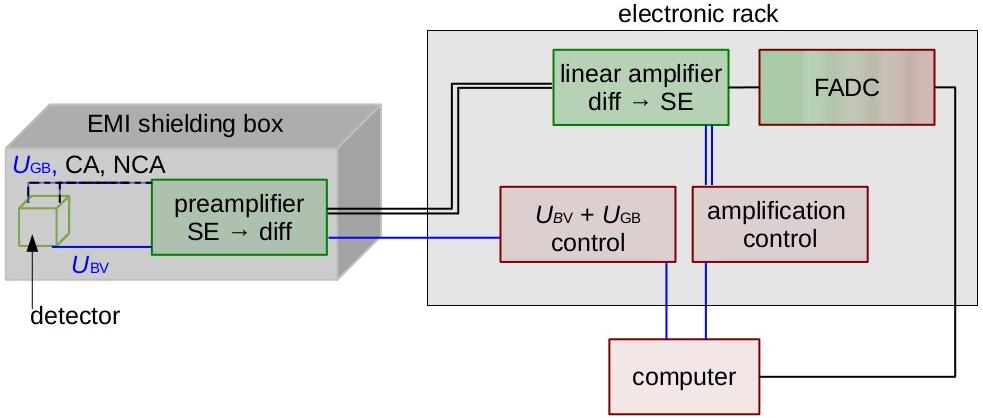}
    \caption[]{The setup used for the presented measurements. The detector, the preamplifier and the radioactive source are placed inside the EMI shielding box, whereas the linear amplifier, the FADC and the control units are located outside the shielding in an electronic rack.}
    \label{fig:daq_chain_schematic}
\end{figure}

The detector as well as the preamplifier are placed inside an electromagnetic interference (EMI) shielding box to shield the detector's unamplified signals from electromagnetic radiation. The calibration sources are placed near the detector inside the shielding box. The remaining readout electronics is installed outside the EMI shielding box to reduce electric and thermal load. The data acquisition chain (DAQ) is shown schematically in Figure \ref{fig:daq_chain_schematic}.

If not stated otherwise, the same operation voltages as used by the manufacturer for the characterization of the detector are applied for the measurements presented in the following, i.e. $U_{\text{GB}}$ = \SI{-80}{\volt} and $U_{\text{BV}}$ = \SI{-1.5}{\kilo\volt}. 
The DAQ system records the pulse shapes of all eight anode electrodes for a time interval of \SI{10.24}{\micro\second} if at least one anode pulse reaches a specified threshold. 

The minimal trigger threshold which can be reached with this setup is \SI{28}{\kilo\electronvolt}. Electronic noise becomes dominant for lower energies depending on the temperature and the humidity in the setup.
For the measurements with a radioactive source, the trigger threshold is set to values higher than the minimal threshold so as to reduce electronic noise. In addition, an analysis threshold is set to values of about \SI{5}{\kilo\electronvolt} higher to operate in a region of constant efficiency. 
The detector specifications and the operating parameters chosen for the majority of the measurements presented in this paper are listed in Table \ref{tab:detector_parameters}.

\begin{table}[h!]
    \begin{center}
    \caption[detector parameters]{Detector specifications and operating parameters.}
    \begin{tabular}{ll}
        \toprule
        Parameter & Value   \\
        \midrule
        Crystal size ($X \times Y \times Z$) & \SI[product-units = power]{2x2x1.5}{\centi\meter} \\
        Crystal mass & \SI{35.2}{\gram} \\
        Single CPG size ($X \times Y$) & \SI[product-units = power]{1x1}{\centi\meter} \\
        CPGs on anode side & 4 \\ 
        Cathode & 1 \\ 
        $U_\text{GB}$ & \SI{-80}{\volt} \\
        $U_\text{BV}$ & \SI{-1.5}{\kilo\volt} \\
        Minimal trigger threshold & \SI{28}{\kilo\electronvolt} \\ 
        \bottomrule
    \end{tabular}
    \label{tab:detector_parameters}
    \end{center}
\end{table} 

\section{Detector characteristics} \label{sec:det_characteristics} 
\subsection{IV characteristics}

The currents between the anodes and the cathode as well as between the anodes within one sector are measured as a function of the applied voltage. Other channels than the measured ones are on a floating potential.
The bulk current between the cathode and the different anodes is measured for bias voltages in a range from \SIrange{0}{-2}{\kilo\volt} at room temperature in steps of \SI{50}{\volt}. The intragrid current between the two electrodes of each sector is measured with a grid bias from  \SIrange{0}{-100}{\volt} in steps of \SI{5}{\volt}. For each applied voltage $U$, 25 single measurements of the resulting current $I$ are performed and the mean values and standard deviations are calculated. 

During the measurements it has been observed that both currents gradually decrease with operation time if the detector has not been operated over longer time periods. Investigations of this behavior are still ongoing. The measurements presented here are started after the currents have been stabilized.  

\begin{figure}[h!]\centering
    \includegraphics[width=0.99\columnwidth, angle=0]{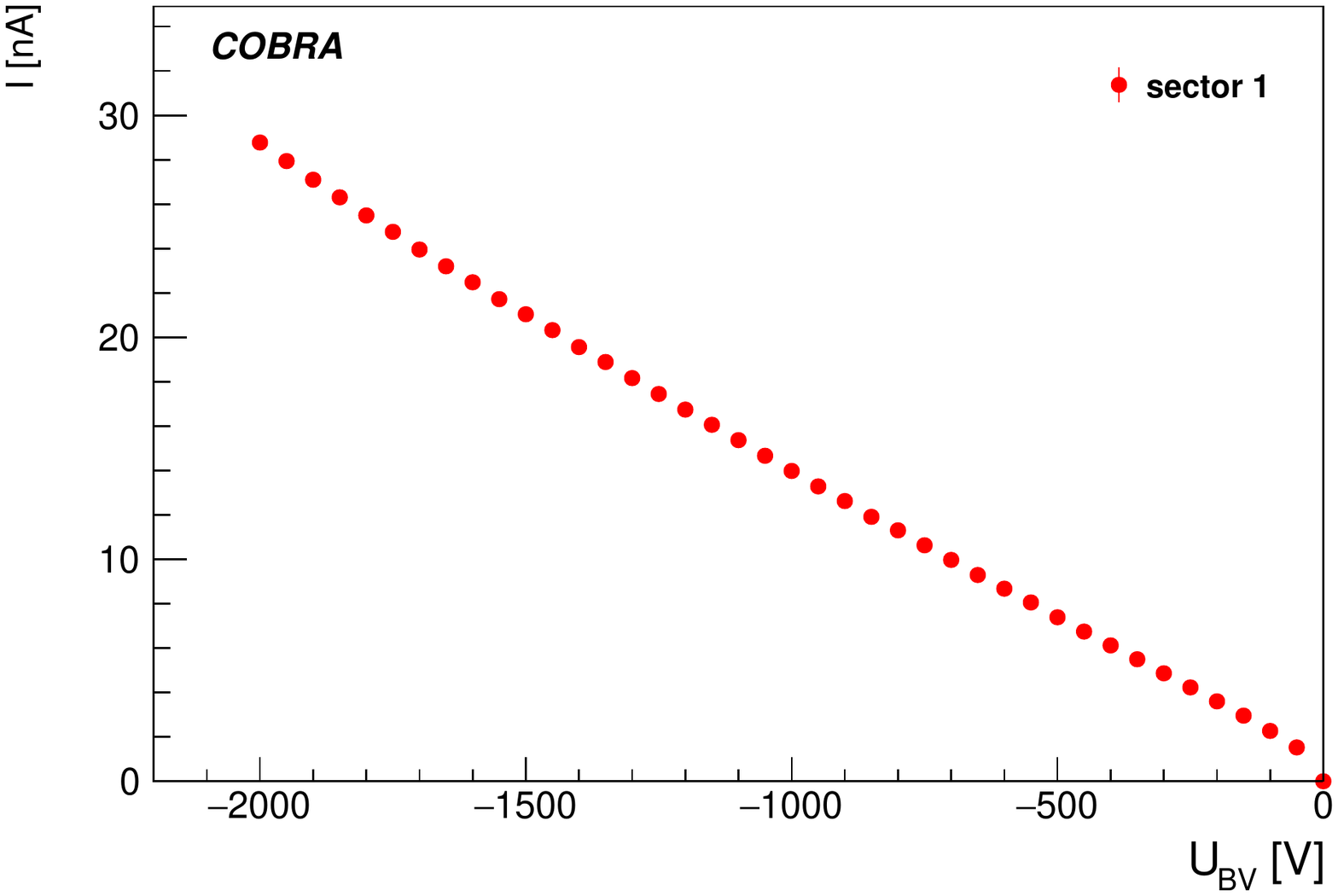}
    \caption[]{The measured current between cathode and the CA of sector one as a function of the bulk voltage. The uncertainties plotted are smaller than the marker size.}
    \label{fig:IV_large_quad_BV}
\end{figure}

Figure \ref{fig:IV_large_quad_BV} shows the averaged bulk current of sector one as a function of the applied voltage. The currents for all sectors rise almost linearly with increasing negative bias voltage up to about \SI{30}{\nano\ampere}.
The maximal difference between the bulk currents of the four sectors is about 5\%. The four curves show the characteristic of an ohmic resistor as expected for a CdZnTe semiconductor detector. The resistance is about \SI{70}{\giga\ohm} for all sectors.

\begin{figure}[h!]\centering
    \includegraphics[width=0.99\columnwidth, angle=0]{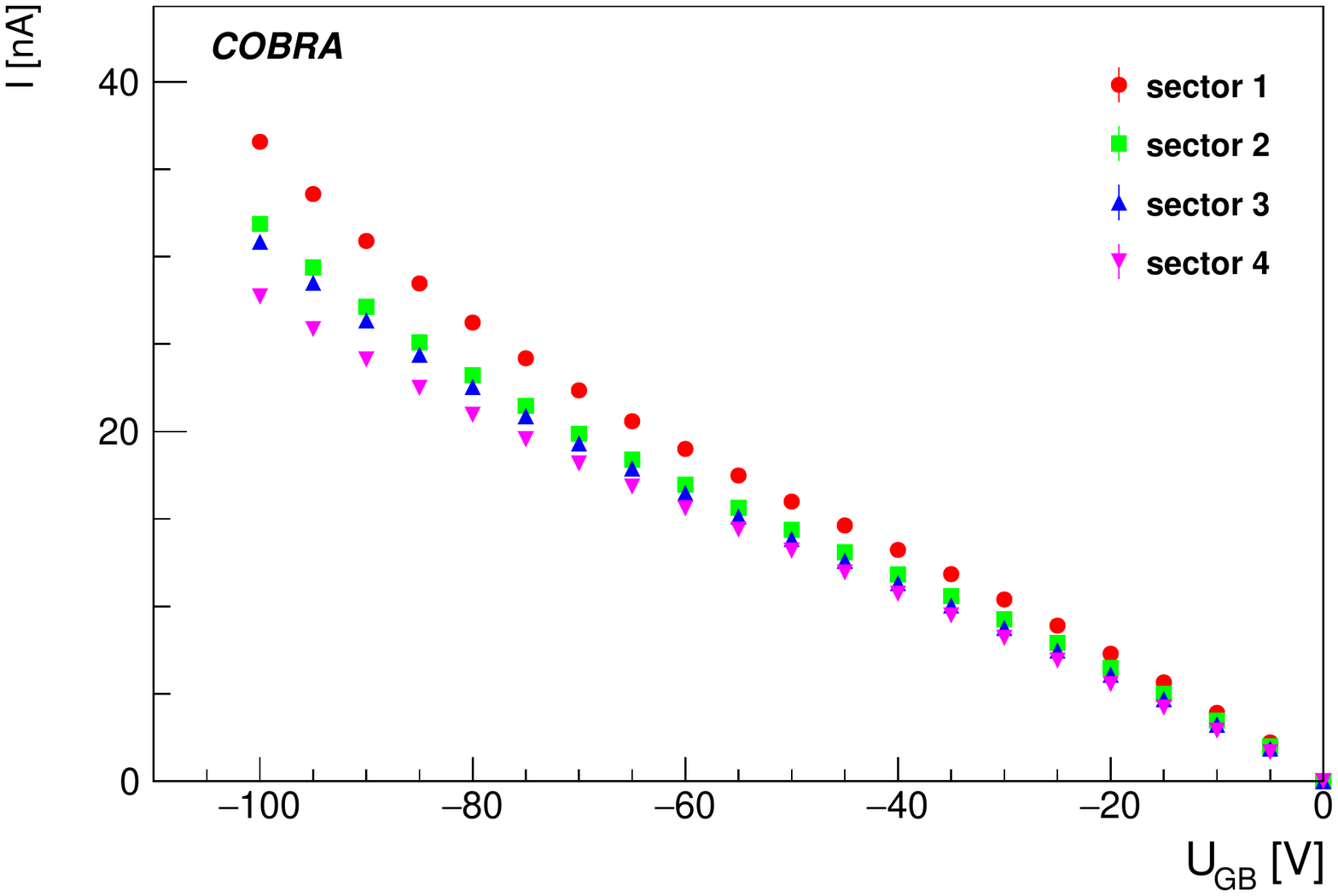}
    \caption[]{The measured currents between the CA and NCA of the four sectors as a function of the grid bias voltage. The uncertainties plotted are smaller than the marker sizes.}
    \label{fig:IV_large_quad_GB}
\end{figure}

Figure \ref{fig:IV_large_quad_GB} shows the intragrid currents between CA and NCA in the four different sectors. The current in sector one rises linearly up to \SI{22}{\nano\ampere} in a range from \SIrange{0}{-70}{\volt} and then increases faster up to about \SI{36}{\nano\ampere} at \SI{-100}{\volt}. Sectors two and three behave similarly to each other. The current rises also linearly but in a wider range from \SIrange{0}{-80}{\volt} before it increases up to approximately \SI{31}{\nano\ampere}. The current of sector four shows a linear behavior as well and rises up to \SI{28}{\nano\ampere}. The curves of all sectors vary maximally by about 28\%. 

At an operating voltage of $U_{\text{BV}} = \SI{-1.5}{\kilo\volt}$ and $U_{\text{GB}} = \SI{-80}{\volt}$, the currents between the two anodes of each sector add up to a total current on the anode side of approximately \SI{90}{\nano\ampere}. The bulk current is about \SI{24}{\nano\ampere} if all anodes are connected.

\subsection{CV characteristics}
Semiconducting devices are also characterized by the relationship between the capacitance $C$ of the electrodes and the applied voltage $U$ between them. 

The applied voltages are in the range from \SIrange{0}{-1.5}{\kilo\volt} in case of measurement of the bulk capacitance and for measurements of the intragrid capacitance they are in a range from \SIrange{0}{-100}{\volt}. The measurements are done with a \textit{HP} 4284A LCR-Meter and a \textit{Keithly} 248 as voltage supply. A custom-made bias-box is used to apply the voltage to the detector without affecting the measurement. The frequency of the applied voltage is \SI{1}{\mega\hertz} for all measurements. 

The bulk capacitance between the cathode and one anode of all four sectors behaves similarly. It is of the order of \SI{0.5}{\pico\farad} and almost constant over the full measurement range up to \SI{-1.5}{\kilo\volt}. The deviations from the mean value are below $1\%$.

The intragrid capacitance between two anodes of the different sectors is roughly \SI{8.0}{\pico\farad} and almost independent of the operation voltage over the full range from \SIrange{0}{-100}{\volt}. The deviations are lower than $0.2\%$.

Neither the bulk nor the intragrid capacitances vary significantly as a function of the applied voltage and therefore behave similarly to an un-doped device as expected for a CdZnTe semiconductor detector.

\section{Energy reconstruction and performance} \label{sec:performance}
In addition to the electric characteristics of the detector, the spectroscopic properties, such as the calibration parameters and the energy resolution, are the most important features for the operation of detectors used for $\gamma$-ray spectroscopy. 
\subsection{Calculation of the deposited energy} \label{subsec:energy_calculation}
As can be shown, the difference between the amplitudes $A$ of the CA and the NCA does not depend on whether the signal of the holes is collected \cite{luke94}. Single-polarity charge sensing is achieved by taking the weighted difference between the two signals. The weight $w$ further compensates for the effect of electron trapping. For details about this technique, see Ref. \cite{fritts_CPG13}. The deposited energy $E$ is thus proportional to  
\begin{align}
    E & \propto A_{CA} - w \cdot A_{NCA} \label{eqn:energy} .
\end{align}

The measured energy in the four sectors is referred to as $E_1$ to $E_4$ and the deposited energy in the whole detector $E_{\text{tot}}$ is calculated by summing up all energies of the single sectors which are above a certain noise threshold.

\begin{figure}[h!]\centering
    \includegraphics[width=0.99\columnwidth, angle=0]{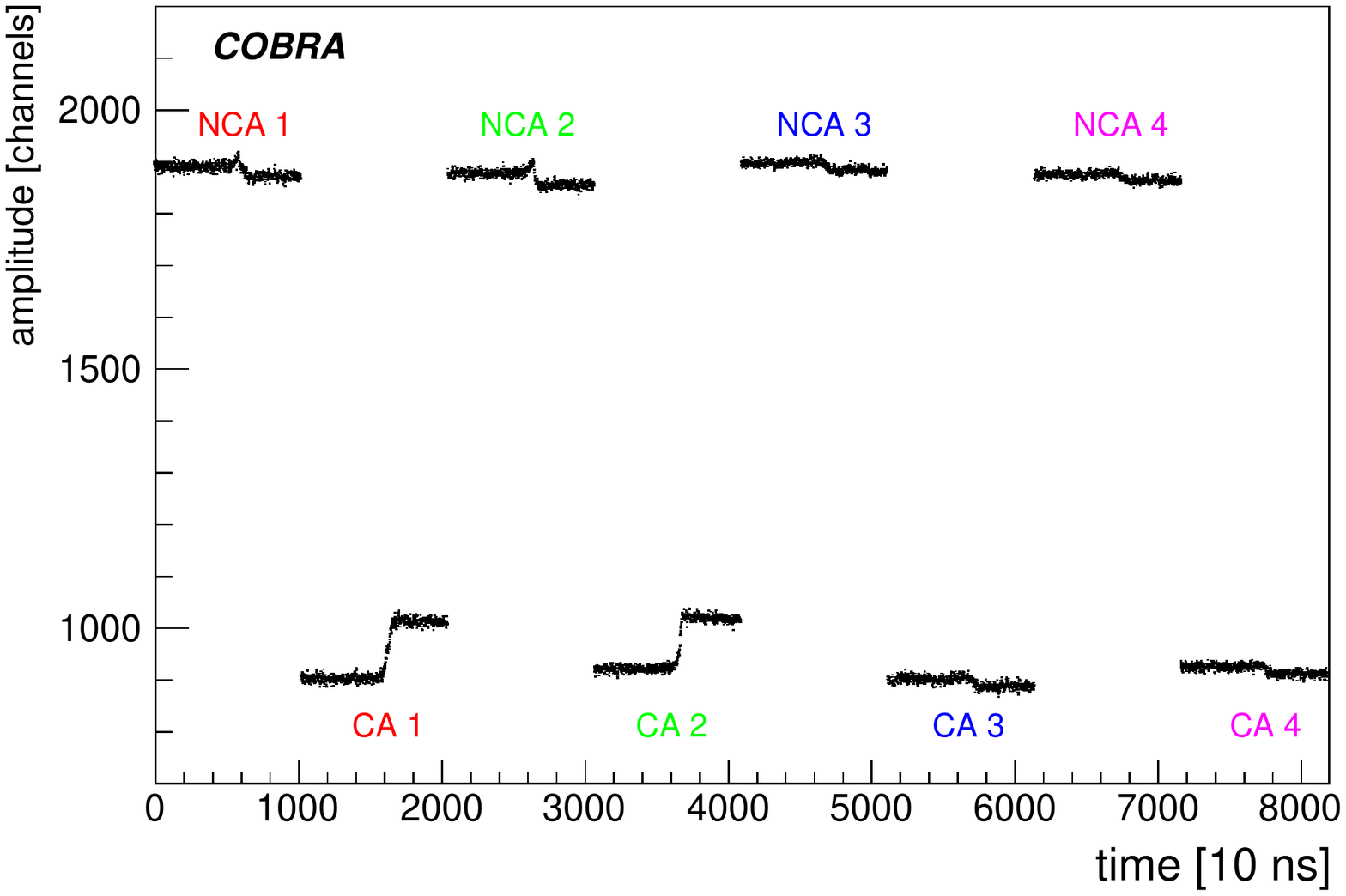}
    \caption[Pulse shape]{Pulse shape of an event depositing \SI{662}{\kilo\electronvolt} from a \isotope[137]{Cs} source in the whole quad-grid detector. Nearly half of the energy is measured in sector one and two, respectively. Sector three and four have no significant energy deposition.} 
    \label{fig:pulse_shape_quad}
\end{figure}

As an example, Figure \ref{fig:pulse_shape_quad} shows the recorded pulse shapes of all eight anodes for a deposited energy of \SI{662}{\kilo\electronvolt} from a \isotope[137]{Cs} source. The eight pulse shapes are shifted by the recorded time interval. The amplitudes of the recorded signals have an arbitrary offset and their heights strongly depend on the adjustment of the linear amplifier. 
Most of the energy is deposited in sector one and sector two, while sectors three and four see mirror charges.

\subsection{Energy calibration and linearity} \label{subsec:energy_response}
The spectroscopic performance of the \SI{6}{\centi\meter\cubed} CdZnTe CPqG detector is investigated in several measurements over an energy range from about \SI{60}{\kilo\electronvolt} to more than \SI{2600}{\kilo\electronvolt}. The various full energy peaks (FEPs) of five employed radioactive sources are listed in Table \ref{tab:several_sources}.

\begin{table}[h!]
    \begin{center}
    \caption[]{Employed sources resulting in characteristic photon energies.}
    \begin{tabular}{Srl}
        \toprule
        \multicolumn{1}{c}{Energy [keV]} & Source & Note \\
        \midrule
        59.54 & \isotope[241]{Am} &  \\
        238.63 & \isotope[232]{Th} & from \isotope[212]{Pb} \\
        338.32 & \isotope[232]{Th} & from \isotope[228]{Ac} \\
        569.70 & \isotope[207]{Bi} & \\
        583.19 & \isotope[232]{Th} & from \isotope[208]{Tl} \\
        661.66 & \isotope[137]{Cs} & \\
        1063.66 & \isotope[207]{Bi} & \\
        1173.23 & \isotope[60]{Co} & \\ 
        1332.49 & \isotope[60]{Co} & \\ 
        1770.23 & \isotope[207]{Bi} & \\
        2614.51 & \isotope[232]{Th} & from \isotope[208]{Tl}\\
        \bottomrule
    \end{tabular}
    \label{tab:several_sources}
    \end{center}
\end{table} 

The five sources differ from each other in dimensions and activities. They all are used as pure $\gamma$-emitters and the $\alpha$- and $\beta$-particles are absorbed in the carrier material of the radioactive isotopes. For the following measurements, all emitted $\gamma$-lines from the \isotope[241]{Am}, \isotope[137]{Cs}, \isotope[60]{Co} and \isotope[207]{Bi} source are used. From the \isotope[232]{Th} source, only those $\gamma$-lines are used, which can be clearly separated from other lines of this source given the resolution of the detector.

\begin{figure}[h!]\centering
    \includegraphics[width=0.99\columnwidth, angle=0]{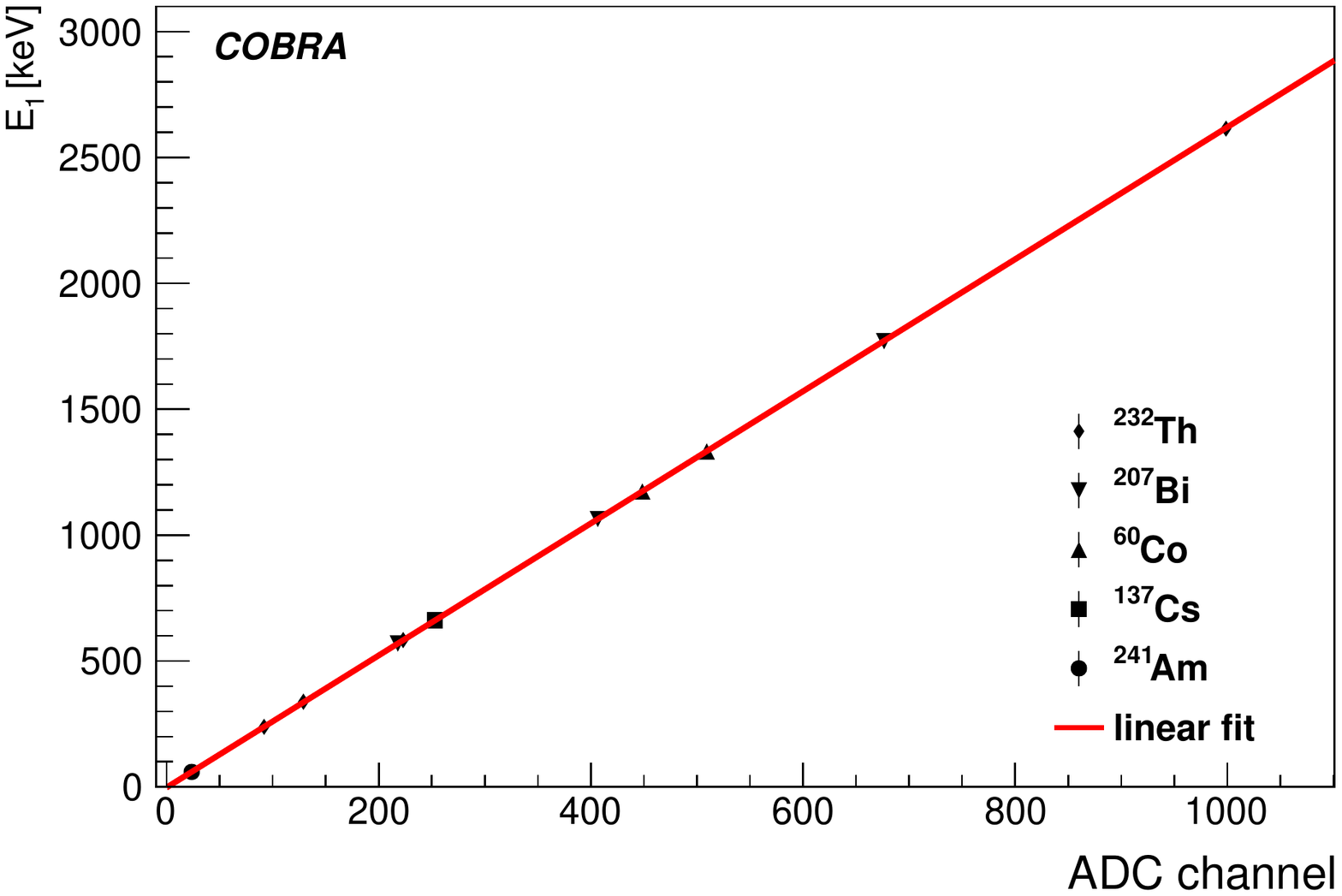}
    \caption[Energy response and residual of four different grids]{The known energies of several $\gamma$-lines are plotted versus the fitted ADC channels to get the energy response of sector one.} 
    \label{fig:energy_response}
\end{figure}

The different $\gamma$-lines in the measured spectra are fitted with a double-Gaussian function plus a linear background model to investigate the energy response and to estimate the energy resolution for various energies. 
Figure \ref{fig:energy_response} shows the known energy of these lines versus the fitted ADC channels.
The linear behavior of the energy response can clearly be seen. 

\begin{figure}[h!]\centering
    \includegraphics[width=0.99\columnwidth, angle=0]{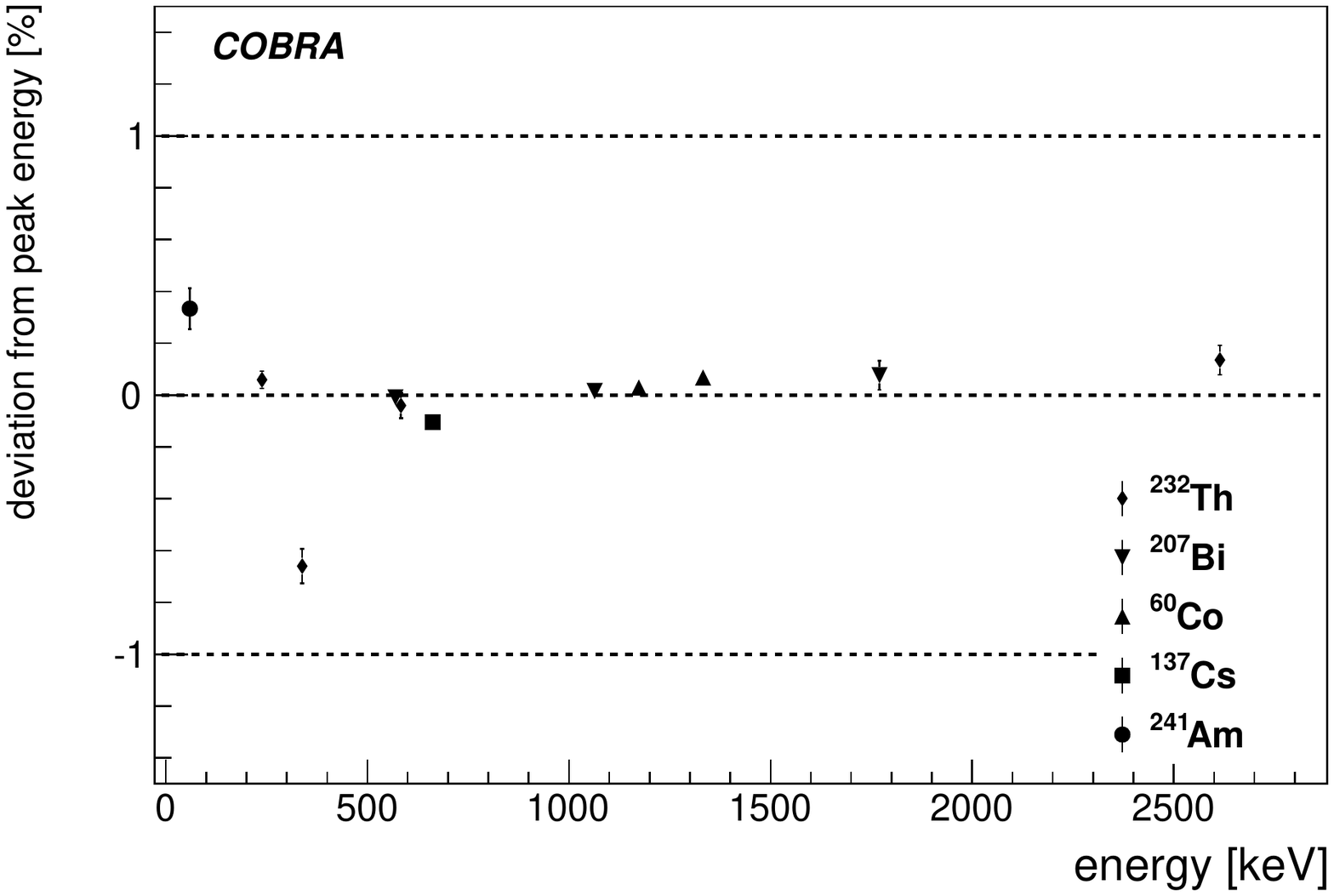}
    \caption[Energy response and residual of four different grids]{The deviations from the known peak energies of sector one are smaller than 1\%.} 
    \label{fig:energy_response_deviations}
\end{figure}

The deviations between the observed energy and the known peak position in the calibrated spectra are shown in Figure \ref{fig:energy_response_deviations}. The four sectors of the quad-grid detector behave very similarly to each other, therefore only one of them is shown in the figures.
The residuals show a spread that is smaller than two percent. The maximum deviations of sectors one to four are 0.64\%, 1.56\%, 0.70\% and 0.44\%, respectively. They are associated with the systematic uncertainty of the calibration of the detector.

\begin{figure}[h!]\centering
    \includegraphics[width=0.99\columnwidth, angle=0]{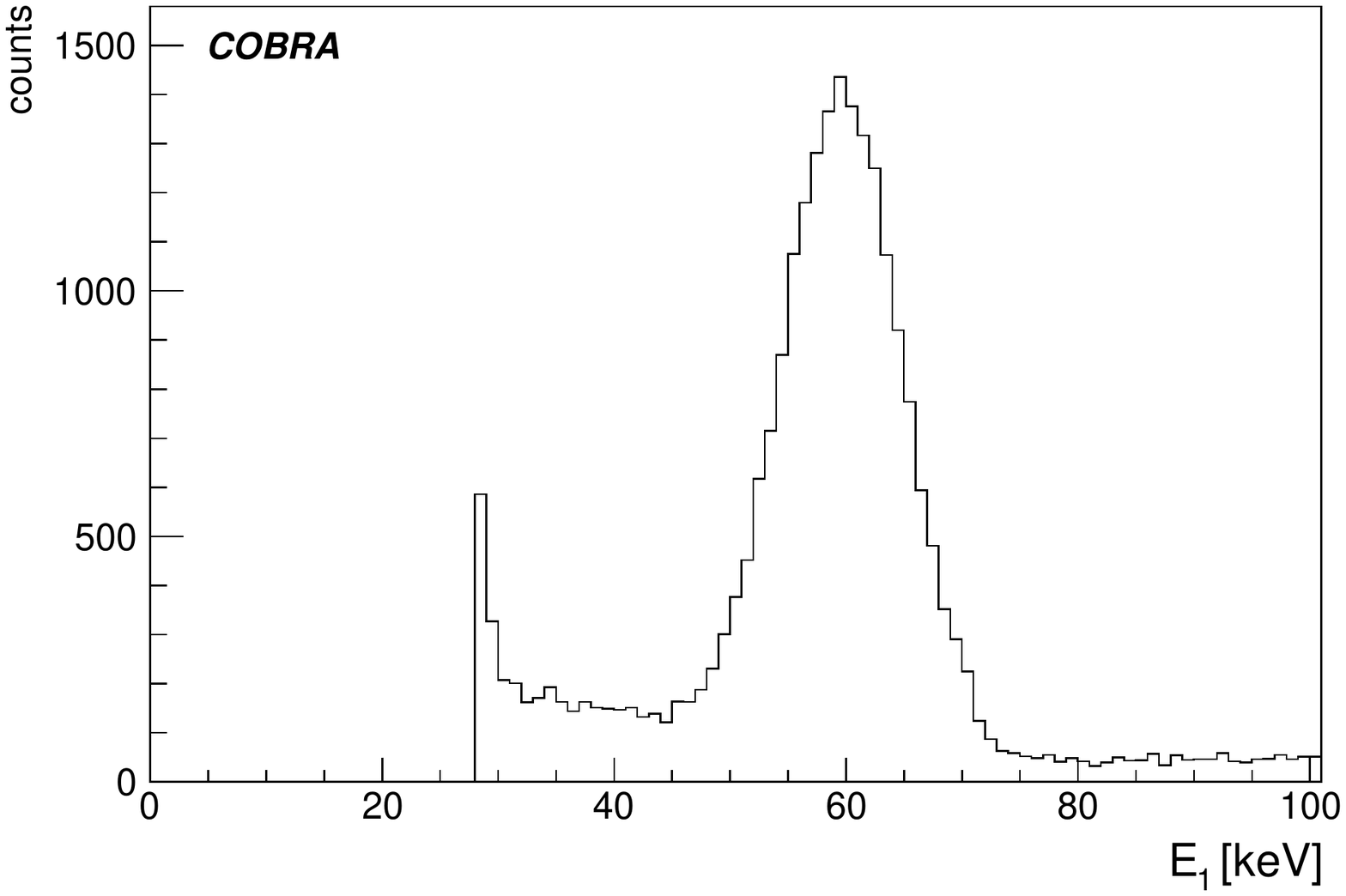}
    \caption[Energy spectra of sector one of different sources]{Energy spectrum of the \isotope[241]{Am} calibration source measured with sector one.}
    \label{fig:several_sources_am}
\end{figure}

\begin{figure}[h!]\centering
    \includegraphics[width=0.99\columnwidth, angle=0]{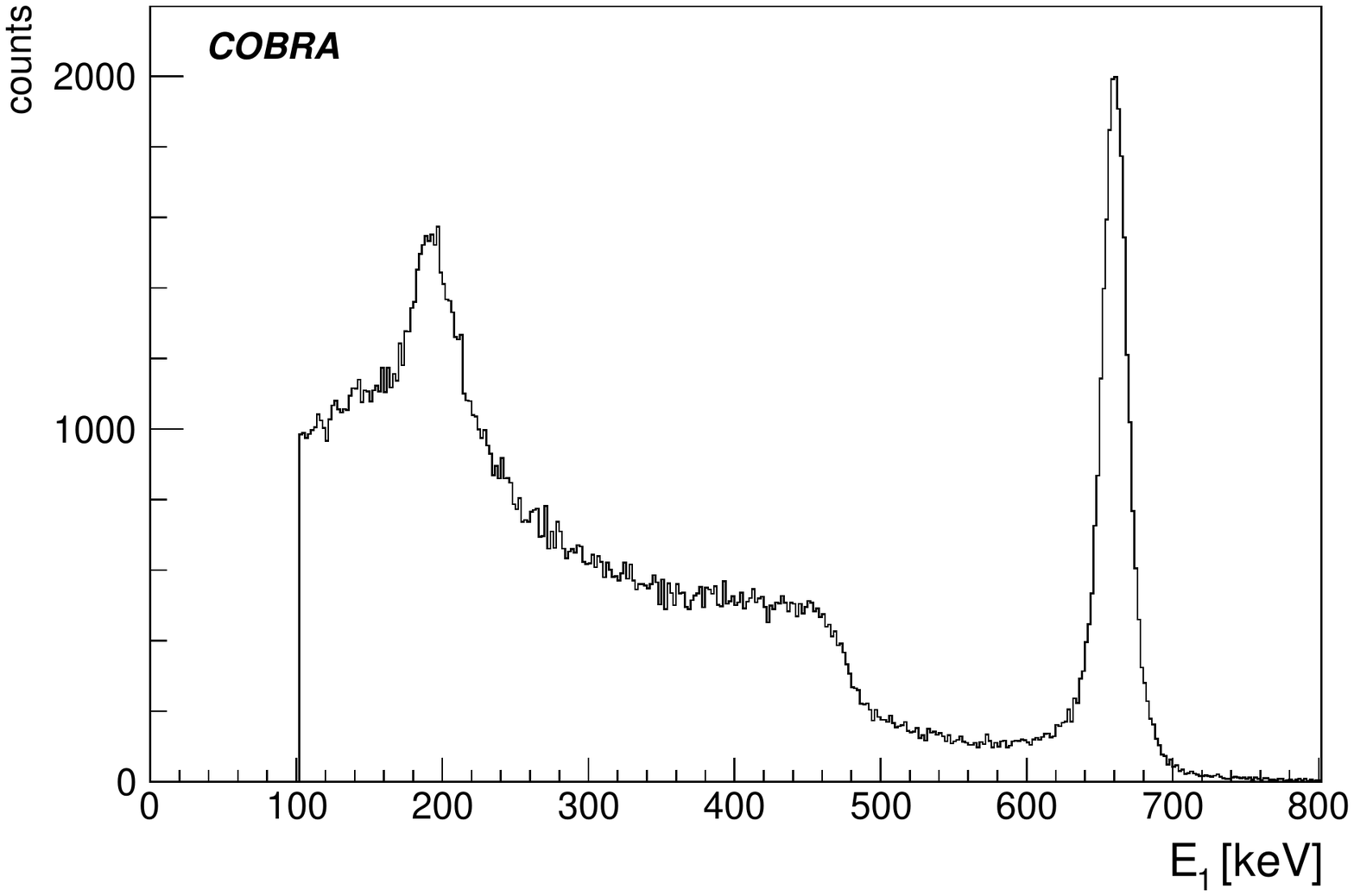}
    \caption[Energy spectra of sector one of different sources]{Energy spectrum of the \isotope[137]{Cs} calibration source measured with sector one.}
    \label{fig:several_sources_cs}
\end{figure}

\begin{figure}[h!]\centering
    \includegraphics[width=0.99\columnwidth, angle=0]{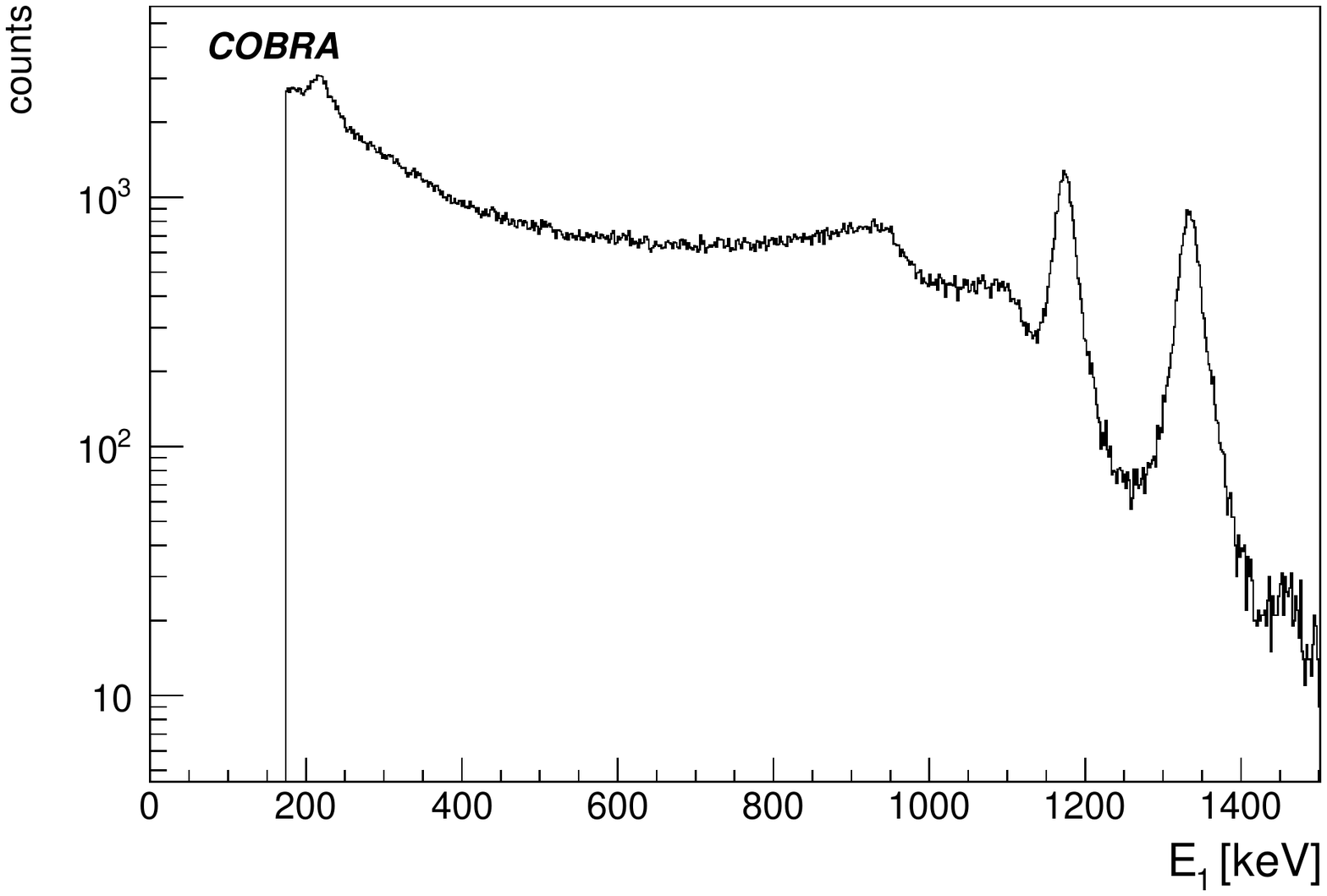}
    \caption[Energy spectra of sector one of different sources]{Energy spectrum of the \isotope[60]{Co} calibration source measured with sector one.}
    \label{fig:several_sources_co}
\end{figure}

\begin{figure}[h!]\centering
    \includegraphics[width=0.99\columnwidth, angle=0]{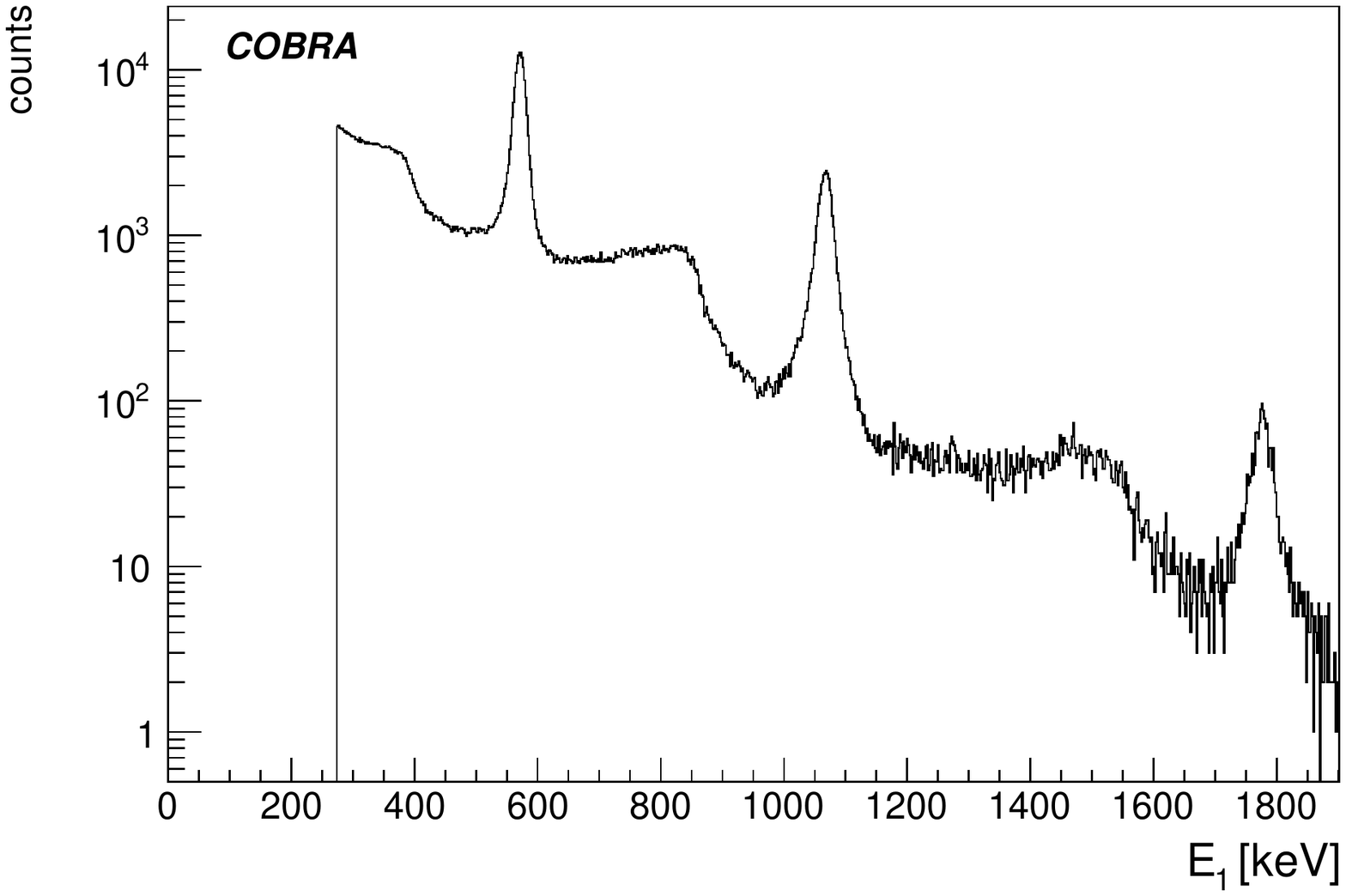}
    \caption[Energy spectra of sector one of different sources]{Energy spectrum the \isotope[207]{Bi} calibration source measured with sector one.}
    \label{fig:several_sources_bi}
\end{figure}

\begin{figure}[h!]\centering
    \includegraphics[width=0.99\columnwidth, angle=0]{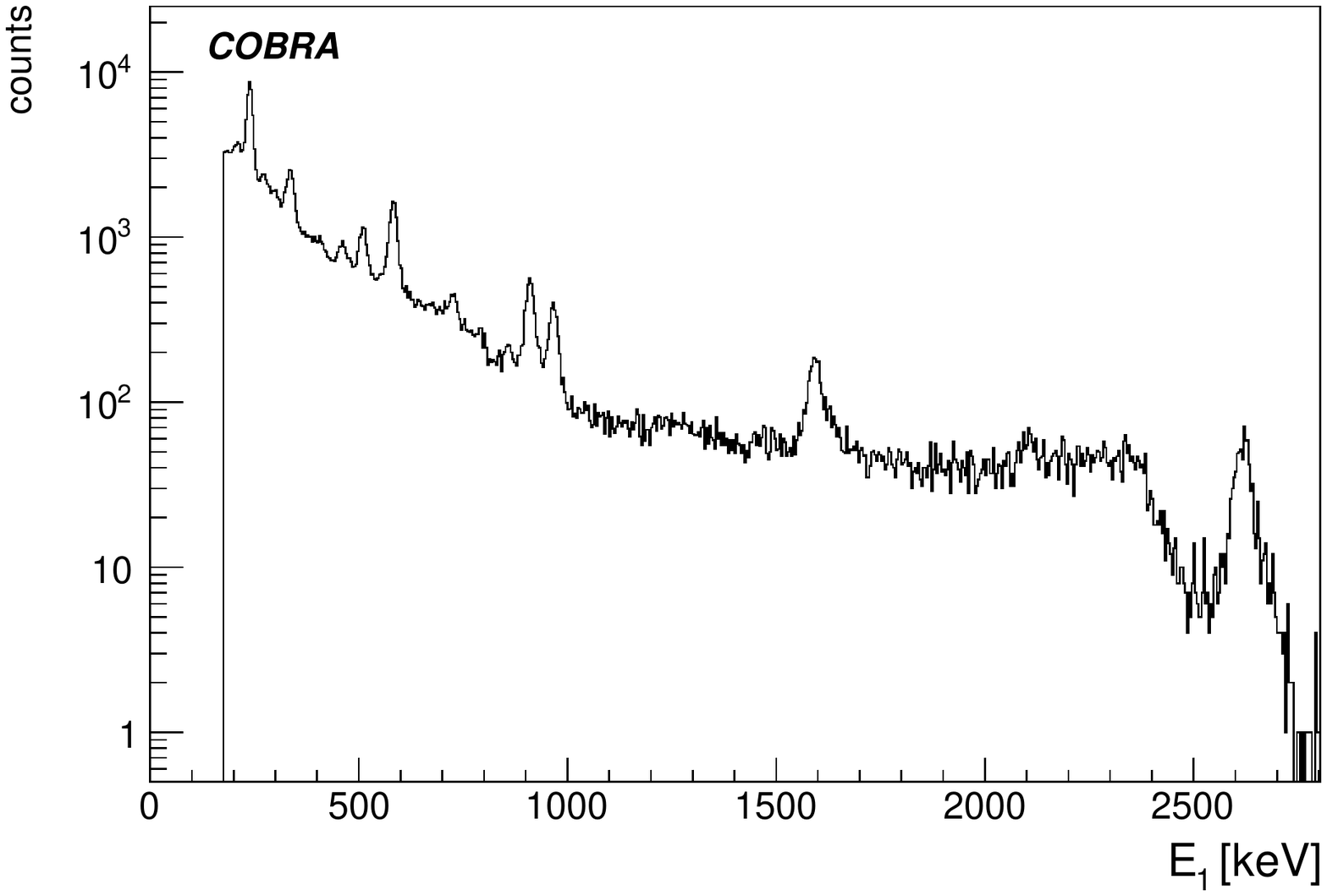}
    \caption[Energy spectra of sector one of different sources]{Energy spectrum of the \isotope[232]{Th} calibration source measured with sector one.}
    \label{fig:several_sources_th}
\end{figure}

\begin{figure}[h!]\centering
    \includegraphics[width=0.99\columnwidth, angle=0]{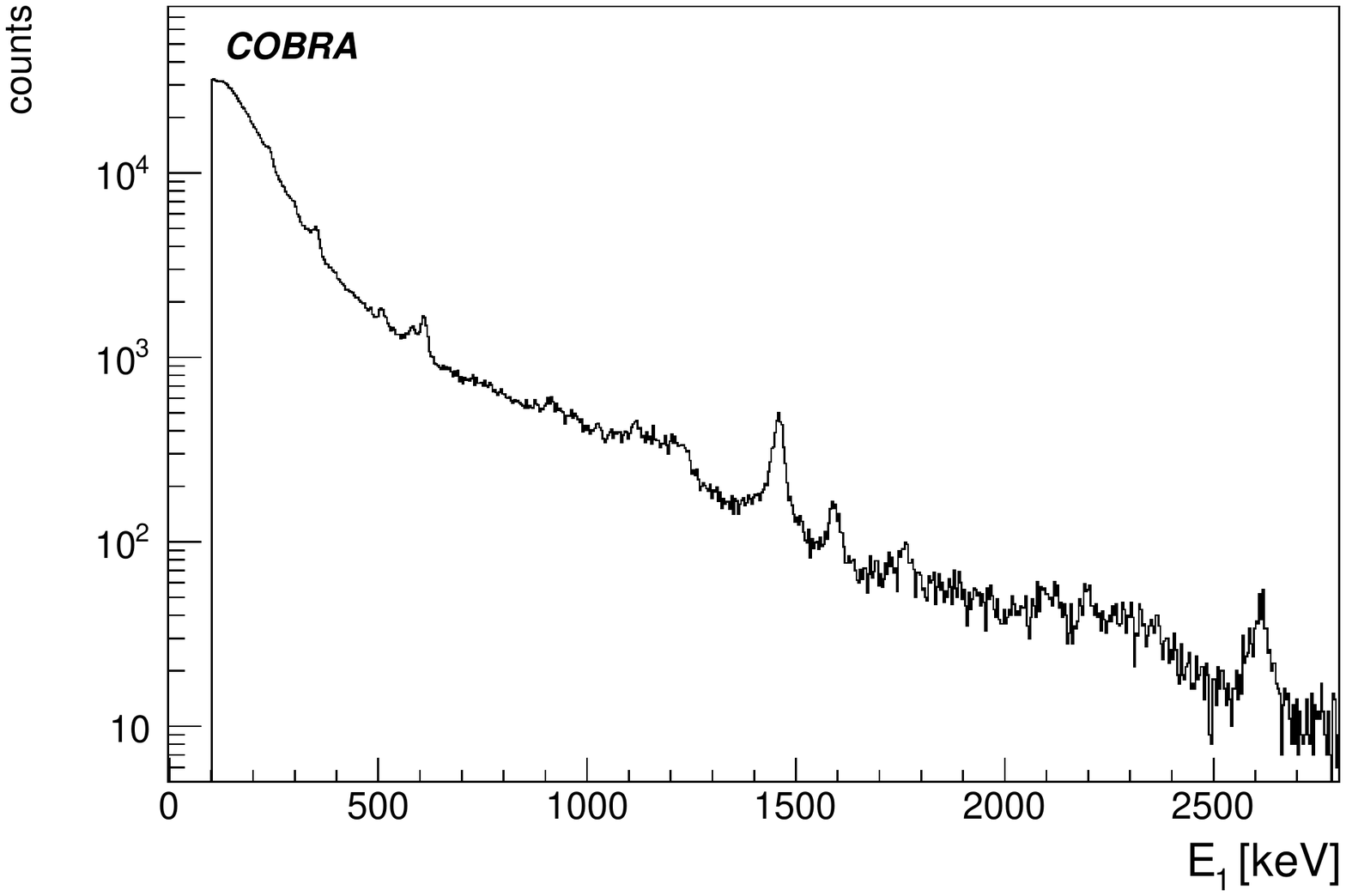}
    \caption[Energy spectra of sector one of different sources]{Energy spectrum of the background in the setup measured with sector one.}
    \label{fig:several_sources_bg}
\end{figure}

The spectra measured with the different sectors are very similar. Thus, Figures \ref{fig:several_sources_am} to \ref{fig:several_sources_th} show the five acquired calibrated energy spectra for sector one. A spectrum from a measurement over one week of the natural radiation in the setup without a calibration source measured with sector one is shown in Figure \ref{fig:several_sources_bg}.

The spectrum of the \isotope[241]{Am} source shows the lowest energetic $\gamma$-line with \SI{59.54}{\kilo\electronvolt} of all used calibration sources (see Fig. \ref{fig:several_sources_am}). For energies lower than \SI{30}{\kilo\electronvolt} the number of events increases due to electronic noise.
This results in the minimal reachable trigger threshold of \SI{28}{\kilo\electronvolt} and the threshold of \SI{33}{\kilo\electronvolt} used in analysis.

\isotope[137]{Cs} has one $\gamma$-line at \SI{662}{\kilo\electronvolt} which can be seen in Figure \ref{fig:several_sources_cs}. The second peak around \SI{185}{\kilo\electronvolt} is the backscatter peak of the $\gamma$-line. Its height depends on the material in the setup and the positioning of the source in the setup.

Figure \ref{fig:several_sources_co} shows the spectrum of the \isotope[60]{Co} source, which emits two $\gamma$-lines at \SIlist{1173;1332}{\kilo\electronvolt}. One of them is located in the Compton valley of the second one. The two corresponding backscatter peaks have an energy of around \SI{212}{\kilo\electronvolt}, but cannot be distinguished due to the energy resolution of the detector.

The three emitted $\gamma$-lines of the \isotope[207]{Bi} source at \SIlist{570;1064;1770}{\kilo\electronvolt} are visible in the spectrum shown in Figure \ref{fig:several_sources_bi}, whereas the corresponding backscatter peaks have energies lower than the trigger threshold employed for this measurement, and therefore they are not recorded.

The spectrum of the \isotope[232]{Th} source (see Fig. \ref{fig:several_sources_th}) features many lines due to the thorium decay chain. For the analysis, only those four lines are used which can be clearly distinguished from neighboring lines. These are the \SI{239}{\kilo\electronvolt} line from \isotope[212]{Pb}, the \SI{338}{\kilo\electronvolt} line from \isotope[228]{Ac} and two lines of \isotope[208]{Tl} (\SIlist{583;2615}{\kilo\electronvolt}).

Figure \ref{fig:several_sources_bg} shows the spectrum of the natural radiation in the setup. The $\gamma$-line of \isotope[40]{K} is clearly visible at \SI{1461}{\kilo\electronvolt} in the spectrum and also other $\gamma$-lines of isotopes from the natural decay chains like \isotope[208]{Tl} (\SIlist{583;2615}{\kilo\electronvolt}), \isotope[212]{Pb} (\SI{239}{\kilo\electronvolt}), \isotope[214]{Bi} (\SIlist{609;1120;1764;2204}{\kilo\electronvolt}), \isotope[214]{Pb} (\SIlist{295;352}{\kilo\electronvolt}) and \isotope[228]{Ac} (\SIlist{911;965;969}{\kilo\electronvolt}) can be identified. 

\subsection{Energy resolution and characteristics} \label{subsec:energy_resolution}
\begin{figure}[h!]\centering
    \includegraphics[width=0.99\columnwidth, angle=0]{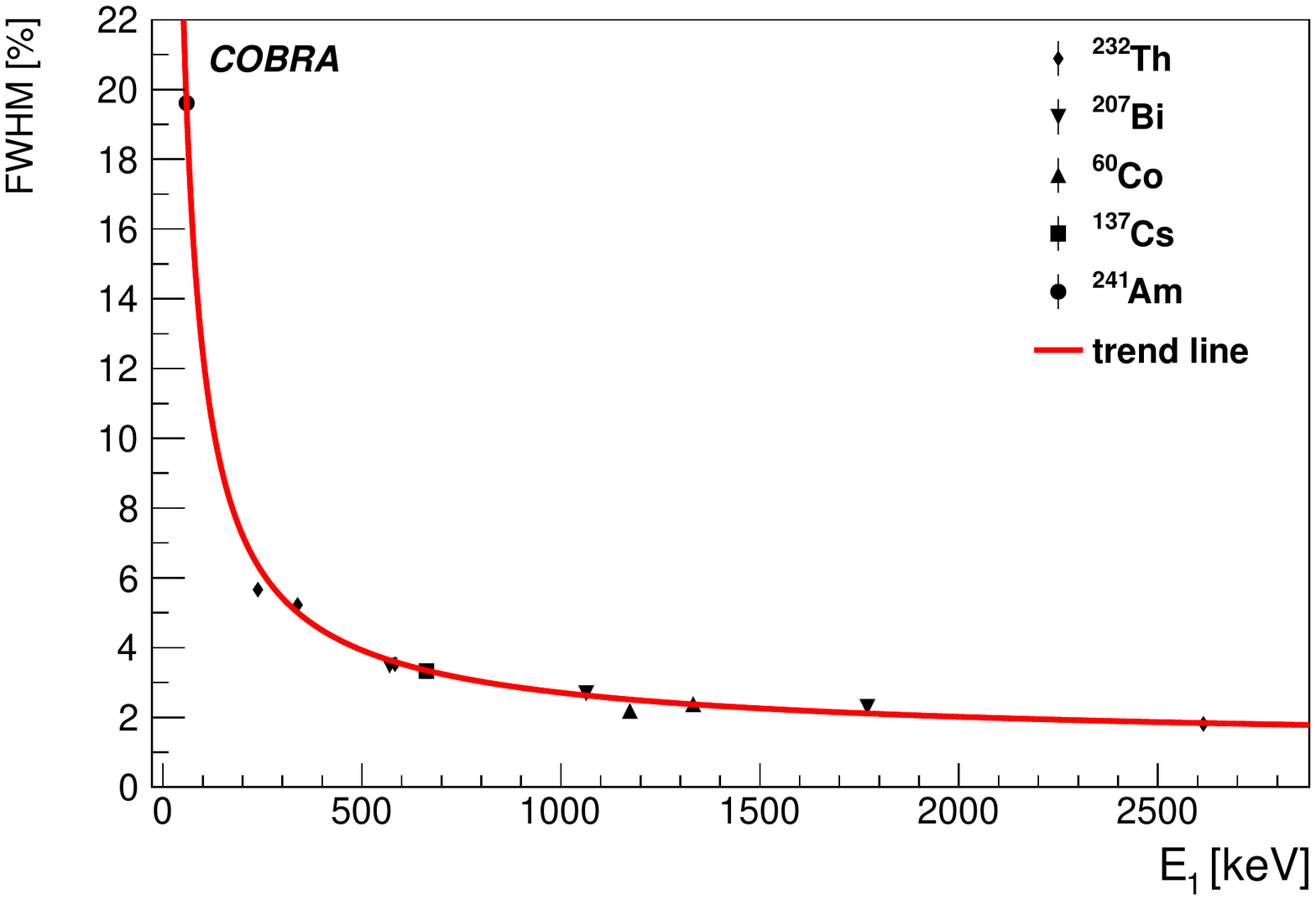}
    \caption[Energy resolution of four different grids]{The FWHM of sector one is measured for several $\gamma$-lines. The trend line is plotted to the energy resolution to guide the eye. The uncertainties plotted are smaller than the marker size.} 
    \label{fig:energy_resolution}
\end{figure}

The calibrated spectra of the sources are used to calculate the energy resolution of the four sectors and the whole detector. Figure \ref{fig:energy_resolution} shows the Full Width at Half Maximum (FWHM) of the different $\gamma$-lines listed in Table \ref{tab:several_sources} measured with sector one. The FWHM decreases from 19.7\% at \SI{60}{\kilo\electronvolt} to 1.8\% at \SI{2615}{\kilo\electronvolt}.

The resolution of the four different sectors behaves similarly and ranges from 3.2\% to 3.8\% at \SI{662}{\kilo\electronvolt} and from 1.7\% to 1.8\% at \SI{2615}{\kilo\electronvolt}. 
The energy resolution of the whole detector is determined the same way as for the single sectors, but with the spectra of $E_{\text{tot}}$ and is calculated to be 2.4\% at \SI{2615}{\kilo\electronvolt} for example.   

Several detector characteristics, such as energy resolution, Full Width at Quarter Maximum (FWQM), photo-fraction and the peak-to-valley (P/V) and peak-to-Compton (P/C) ratio, are measured with a \isotope[137]{Cs} calibration source. These are listed in Table \ref{tab:characterizing_parameters}.

\begin{table*}
    \begin{minipage}[c]{0.99\textwidth} 
        \begin{center}
        \caption[]{Several detector parameters calculated at the full energy peak of a \isotope[137]{Cs} source.}
            \begin{tabular}{llllll}
                \toprule
                Parameter & \multirow{2}{*}{Sector 1} & \multirow{2}{*}{Sector 2} & \multirow{2}{*}{Sector 3} & \multirow{2}{*}{Sector 4} & \multirow{2}{*}{Detector} \\ 
                @ \SI{662}{\kilo\electronvolt}   & & & & & \\
                \midrule
                FWHM [\%]                              & \num{3.33\pm0.08}    & \num{3.25\pm0.09}    & \num{3.76\pm0.05}    & \num{3.48\pm0.03}    & \num{3.84\pm0.02} \\
                FWHM [keV]                              & \num{22.0\pm0.5}    & \num{21.5\pm0.6}    & \num{24.9\pm0.3}    & \num{23.0\pm0.2}    & \num{25.4\pm0.1} \\
                FWQM [keV]                              & \num{32.5\pm0.7}    & \num{31.5\pm1.1}    & \num{39.3\pm0.5}    & \num{33.8\pm0.4}    & \num{37.6\pm0.2} \\
                FWQM/FWHM                               & \num{1.47\pm0.05}      & \num{1.47\pm0.07}      & \num{1.58\pm0.03}      & \num{1.47\pm0.02}      & \num{1.48\pm0.01}  \\
                P/V                                     & \num{13.4\pm0.2}     & \num{14.7\pm0.3 }    & \num{8.7\pm0.1}     & \num{10.6\pm0.2}     & \num{11.3\pm0.1} \\ 
                P/C                                     & \num{3.7\pm0.03}      & \num{3.81\pm0.03}      & \num{2.92\pm0.03}     & \num{3.27\pm0.03}      & \num{4.57\pm0.02} \\ 
                photo-frac. [\%]                      & \num{14.0\pm0.1}    & \num{14.2\pm0.1}    & \num{14.4\pm0.1}   & \num{13.7\pm0.1}    & \num{19.6\pm0.1} \\
                \bottomrule
            \end{tabular}
        \label{tab:characterizing_parameters}
        \end{center}
    \end{minipage}
\end{table*}

As expected, the whole detector shows a higher peak-to-Compton ratio and a larger photo-fraction than the single sectors because of the higher probability to deposit the total energy of an incoming photon in the whole detector rather than in a single sector. A further discussion of this effect can be found in Section \ref{sec:grid_correlation}.

\section{Monte Carlo simulations} \label{sec:simulations} 
The experimental results are compared to Monte Carlo simulations. The simulations are performed using the COBRA Monte Carlo framework VENOM, which is based on the GEANT4 framework \cite{Agostinelli2003250} and is tailored to low-background applications such as COBRA. 

The implemented detector geometry includes the detector crystal, the electrodes and the coating of the crystal, as well as the detector holding structure and the enclosure of the radioactive sources. High-density and high-Z materials near the source and the detector are also considered in the simulation. The \isotope[241]{Am}, \isotope[137]{Cs}, \isotope[60]{Co}, \isotope[207]{Bi} and \isotope[232]{Th} source have been simulated. The position of these sources in the real setup are measured and replicated in the simulation.

As VENOM does not take into account the detector response, the energy resolution is convolved into the simulated data in a second step. This is done by adding a random term to each energy deposition. The random term is drawn from a double-Gaussian distribution for which the parameters are obtained by a fit to the data. 

To account for signals from the natural radiation background in the laboratory, events from a long-term measurement of the background are randomly added to the simulated spectra. As the exact activity of the sources has not been measured, the relative contribution from background radiation has been determined by a fit to the height of the peak of \isotope[40]{K} at \SI{1461}{\kilo\electronvolt}. To suppress the effects of events triggered by 
electronic noise, only events above \SI{350}{\kilo\electronvolt} are considered. The \isotope[241]{Am} source, which produces no $\gamma$-rays with significant intensity above \SI{60}{\kilo\electronvolt}, is simulated without a background contribution because of its high activity.

\begin{figure}[h!]
    \centering
    \includegraphics[width=0.99\columnwidth]{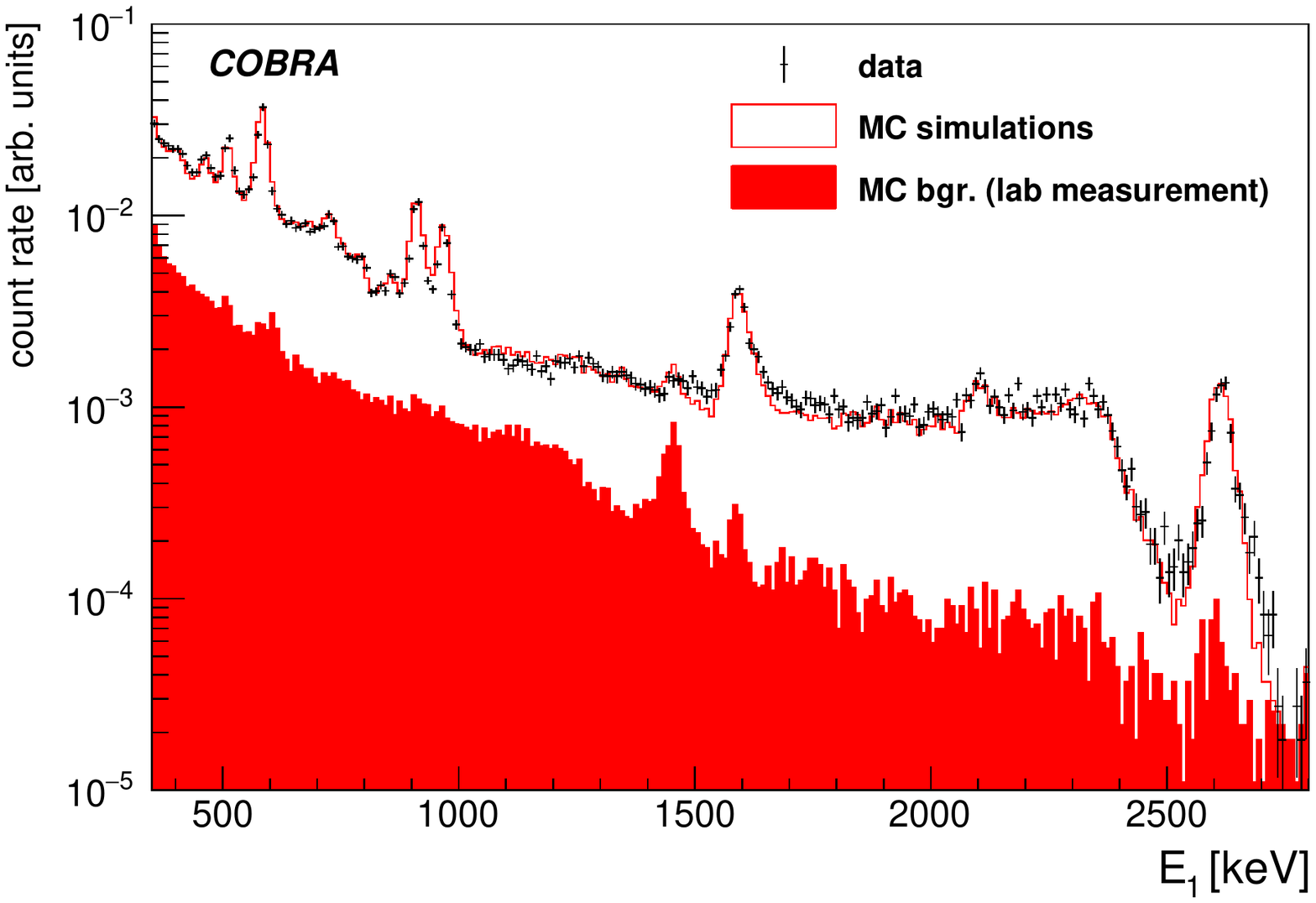}
    \caption{Comparison of Monte Carlo (solid red line) and measurement (black points) spectrum of the \isotope[232]{Th} source. The background contribution to the Monte Carlo is shown in the solid red area.}
    \label{fig:Spectrum_232th}
\end{figure}

As an example, the simulated \isotope[232]{Th} spectrum is compared to the measured spectrum and is shown in Figure \ref{fig:Spectrum_232th} for a single sector. Except for minor differences in the lineshapes, no systematic deviations between the simulated and the measured data are observed.  

\section{Operational stability}\label{sec:stability}
The operational stability of the detector is tested by measurements of the count rate, the peak position and energy resolution over a period of several days.
The development of the stability is analyzed by fitting a linear function to the data points of each measurement.

\begin{figure}[!ht]\centering
    \includegraphics[width=0.99\columnwidth, angle=0]{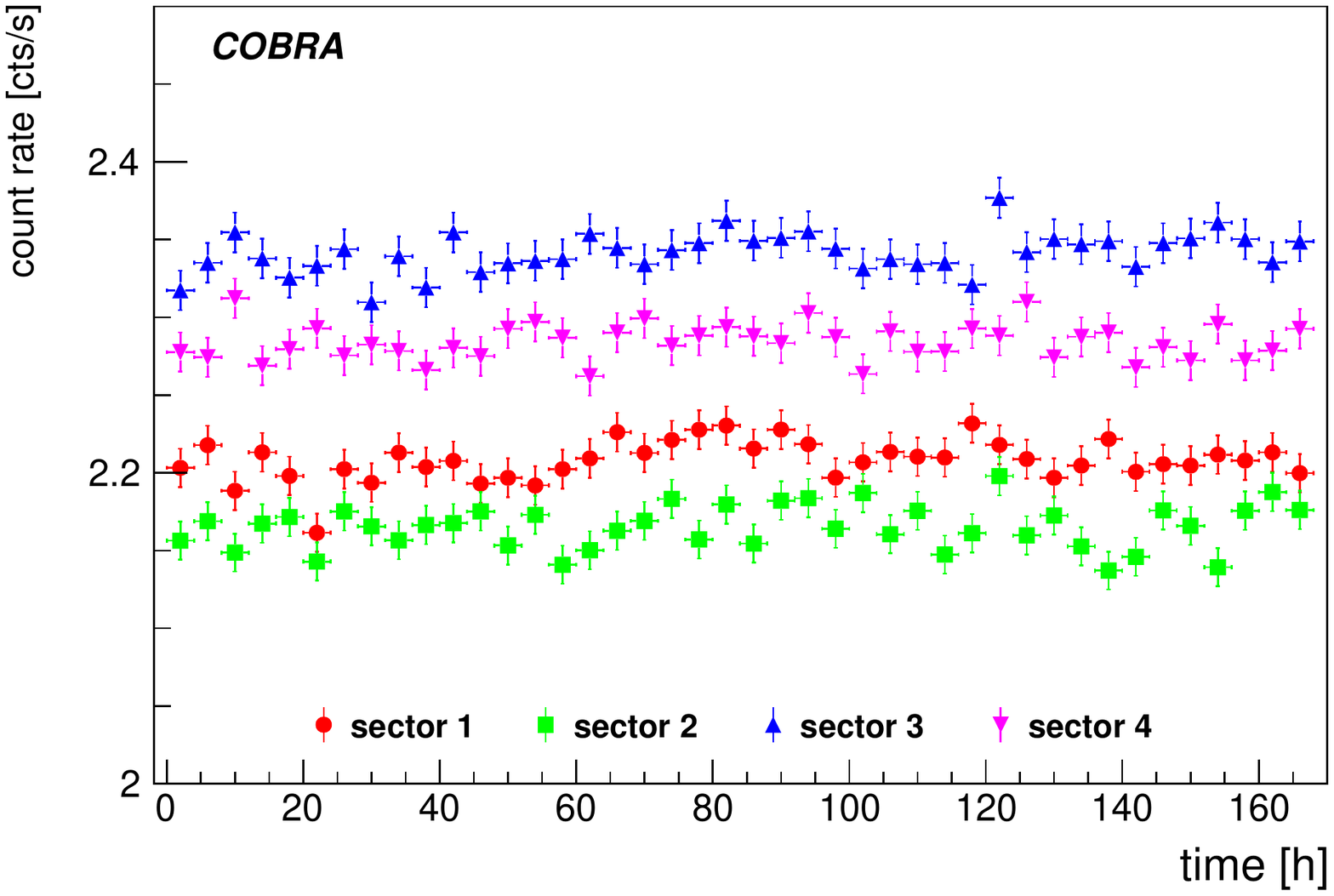}
    \caption[Operational stability of background count rates]{Stability of count rates from background for one week measurement time.}
    \label{fig:stability_countrates}
\end{figure}

The stability of the count rate is tested with a background measurement over one week. Figure \ref{fig:stability_countrates} shows the constancy of the count rate for every sector in a window from \SIrange{90}{2700}{\kilo\electronvolt} over time. The average count rate of the four sectors ranges from \SI{2.16}{counts\per\second} to \SI{2.34}{counts\per\second}.
A slightly different environment of the natural radiation in the setup for the four sectors could cause the small differences between the count rates.
The slopes of the fitted functions agree with zero within 2.5 standard deviations.

To test the stability of the energy response, a measurement over one week with a \isotope[137]{Cs} source at room temperature is performed. Every two hours a spectrum is taken for 10 minutes. The calibration parameters are determined for the first measurement and during the whole measurement no re-calibration is done. 
The change of the peak position of the \SI{662}{\kilo\electronvolt} $\gamma$-line in the spectra is smaller than 0.1\% over the whole measurement period for all sectors. Small fluctuations range from \SIrange{\pm0.6}{\pm0.9}{\kilo\electronvolt} for the different sectors, whereas the energy resolution is of the order of \SI{23}{\kilo\electronvolt} FWHM at this energy range.
Thus, the variation of the peak position is negligible.

The stability of the energy resolution is investigated with the same data. The FWHM of the FEP is calculated as described in Section \ref{sec:performance}.
The energy resolution of every sector itself shows a constant behavior. However, the resolution of every sector fluctuates around slightly different values. 
The fluctuations of the energy resolution within one sector are on average \SI{\pm1}{\kilo\electronvolt}.
It can thus be concluded that the energy response and the resolution of all four sectors, and therefore of the whole detector, is stable during the operation over a period of one week.

Investigations on long-term stability over several weeks of such detectors are ongoing.

\section{Correlation between sectors} \label{sec:grid_correlation}
There are several reasons for coincidences between sectors. First of all, it is possible to measure random coincidences during the \SI{10.24}{\micro\second} recording time of the FADC. The rate of such coincidences is small if the activity of the sources is low enough. This requirement is largely fulfilled by the used sources. In a low-background environment, random coincidences are negligible.
In contrast, the summation peak of the \SIlist{1173;1332}{\kilo\electronvolt} photon from a \isotope[60]{Co} source shows a true physical coincidence. 
This means that two $\gamma$-particles are emitted quasi-simultaneously. As each particle is emitted into an almost random direction, the angular correlation can be neglected for the setup, because only a small part of these physical coincidences can be detected in the detector.

A second class of events is related to the scattering of the incoming radiation in the detector itself. While $\alpha$- and $\beta$-particles have an almost single-site energy deposition, $\gamma$-rays can scatter multiple times within the detector due to Compton-scattering. As the distance between two scattering locations for $\gamma$-rays can be quite large even in materials like CdZnTe, Compton-scattering will often happen across more than one sector. Thus, the identification of events in several sectors can be used to distinguish between single-site and multi-site events.

Because of different electronic effects it is also possible for a single localized event to appear as a multi-sector event. For example, the generated electron cloud is not guaranteed to be contained within its sector of origin. A fraction of the total charge can also drift into another sector. Also, a single-site event can look like a multi-sector event if the crosstalk between different readout channels is not negligible. 
It is expected that both effects are vanishingly small due to the way of operating the four outer grids as CAs and the inner ones as NCAs and the readout electronics used.

As before, the information from all four sectors is read out as soon as the pulse height of at least one anode is above the trigger threshold. Only a very small fraction of events deposits energy in all sectors simultaneously. Most of the time, the pulses of one or more readout channels contain only random noise. Therefore, a naive combination of the measured energies of all four sectors would lead to a spectrum with a significantly worse energy resolution. To avoid this, pulses below the noise threshold are rejected in the analysis presented in this paper.

\subsection{Intergrid spectra correlation}
A very basic way to quantify the influence of multi-sector events is the sector multiplicity. The sector multiplicity is defined as the number of sectors $N$ with an energy deposition above a certain threshold within one event. $N$ is one if the whole measured energy is only localized in one sector and four if all four sectors measure an energy deposition. The consideration of the sector multiplicity shows that regardless of the source, approximately 10\% of the events are multi-sector events. 

\begin{figure}[!ht]\centering
    \includegraphics[width=0.99\columnwidth, angle=0]{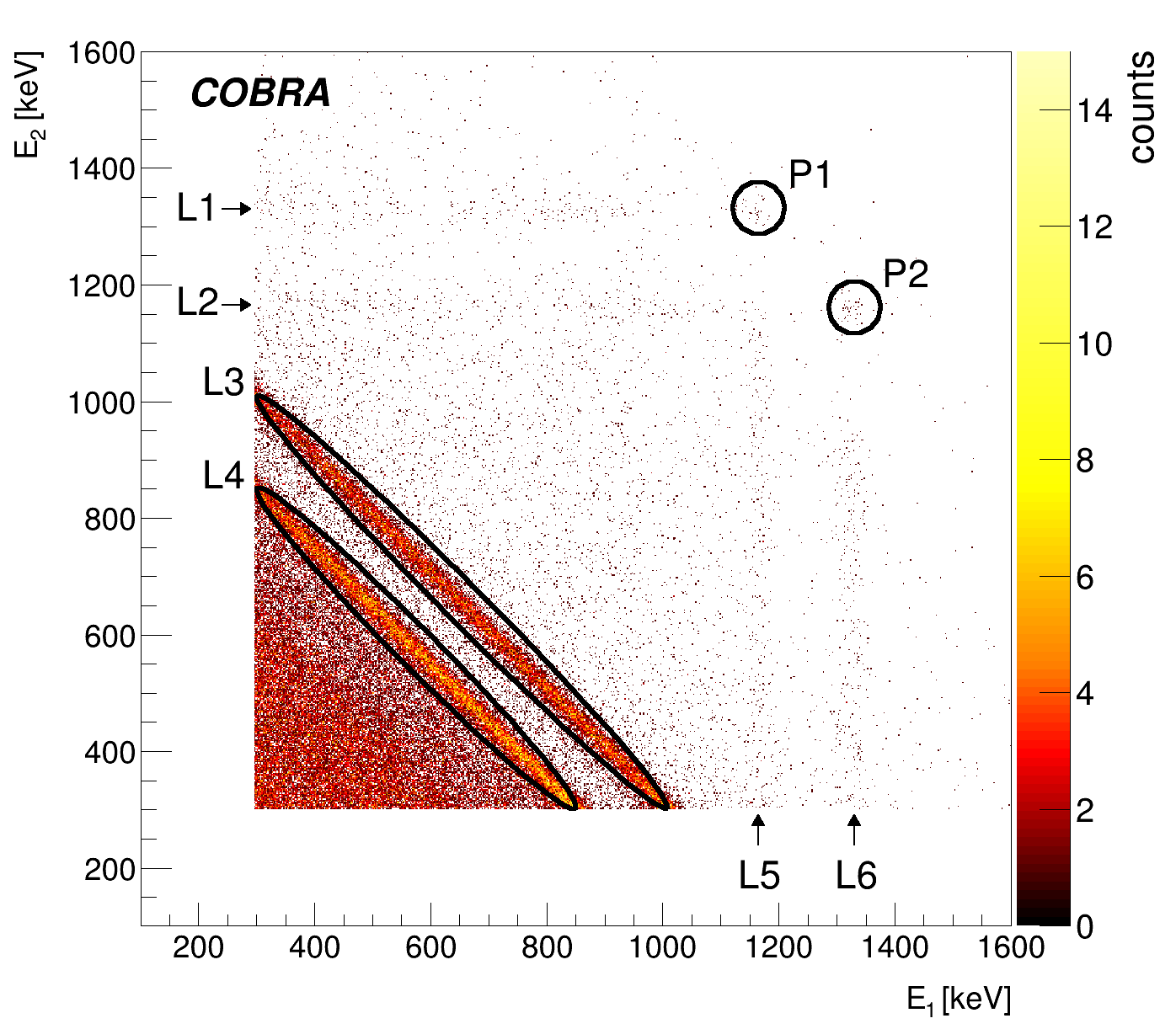}
    \caption[Intergrid correlation]{Correlation between sector one and two for a measurement with a \isotope[60]{Co} source. The simultaneous energy deposition of both photons is clearly visible.}
    \label{fig:intergrid_correlation}
\end{figure}

To further investigate these correlations, it is essential to look at the relationship between the different sectors for those particular events. The analysis here is restricted to two-sector events. Figure \ref{fig:intergrid_correlation} exemplifies the correlation of the deposited energy between the sectors one and two for a \isotope[60]{Co} measurement. This isotope emits two $\gamma$-lines (\SIlist{1173;1332}{\kilo\electronvolt}). They occur both with a probability of more than 99\%. The physical coincidence causes a summation peak at \SI{2505}{\kilo\electronvolt}, whose measured intensity depends only on the solid angle covered by the detector. Due to the large detector volume, it is probable that the two photons interact in more than one sector. This feature and further ones are clearly visible in Figure \ref{fig:intergrid_correlation} and described below.

\begin{description}
    \item[L1:] The full energy of the \SI{1332}{\kilo\electronvolt} photon is deposited in sector two and some additional energy from a Compton-scattered second \SI{1173}{\kilo\electronvolt} photon in sector one.
    \item[L2:] The full energy of the \SI{1173}{\kilo\electronvolt} photon is deposited in sector two and some additional energy from a Compton-scattered second \SI{1332}{\kilo\electronvolt} photon in sector one.
    \item[L3:] The full energy of the \SI{1332}{\kilo\electronvolt} photon is deposited in sector one and two.
    \item[L4:] The full energy of the \SI{1173}{\kilo\electronvolt} photon is deposited in sector one and two.
    \item[L5:] The full energy of the \SI{1173}{\kilo\electronvolt} photon is deposited in sector one and some additional energy from a Compton-scattered second \SI{1332}{\kilo\electronvolt} photon in sector two.
    \item[L6:] The full energy of the \SI{1332}{\kilo\electronvolt} photon is deposited in sector one and some additional energy from a Compton-scattered second \SI{1173}{\kilo\electronvolt} photon in sector two.
    \item[P1:] The full energy of the \SI{1173}{\kilo\electronvolt} photon is deposited in sector one and the full energy of the \SI{1332}{\kilo\electronvolt} photon in sector two.
    \item[P2:] The full energy of the \SI{1332}{\kilo\electronvolt} photon is deposited in sector one and the full energy of the \SI{1173}{\kilo\electronvolt} photon in sector two.
\end{description}

The points P1 and P2 contain those events which belong to the summation peak of \isotope[60]{Co} in the spectrum of the whole detector.

\begin{figure}[!ht]\centering
    \includegraphics[width=0.99\columnwidth, angle=0]{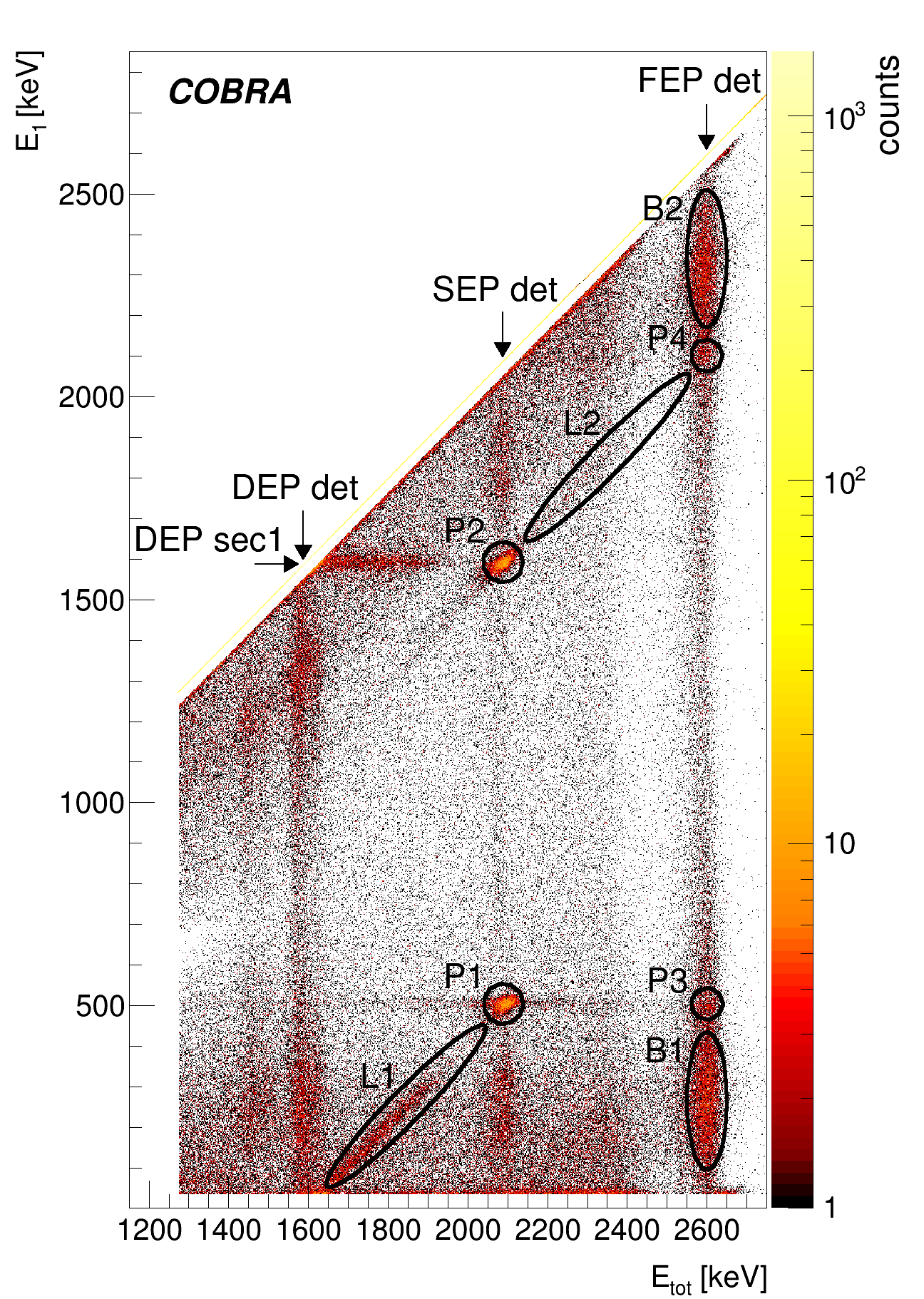}
    \caption[Detector sector correlation]{Correlation between the deposited energy in sector one and in the whole detector for a measurement with a \isotope[232]{Th} source. Most of the different visible features stem from the pair production from a \SI{2615}{\kilo\electronvolt} photon of \isotope[208]{Tl}.}
    \label{fig:detector_grid_correlation}
\end{figure}

Another important correlation is the one between the deposited energy in one sector and the energy deposited in the whole detector. Figure \ref{fig:detector_grid_correlation} shows such a correlation in a measurement with a \isotope[232]{Th} source. The energy spectrum of the source contains multiple $\gamma$-lines, which are visible in the figure. The vertical lines are caused by photons which deposit their whole energy in the detector and only a part of it in sector one. Those are multi-sector events. In contrast, the events on the angle bisector are single-sector events. They deposit their whole energy only in sector one.

Additionally, some events from well known physical processes, particularly induced by the \SI{2615}{\kilo\electronvolt} FEP of \isotope[208]{Tl}, can be identified in the scatter plot. Photons in this energy region have a relative high cross section for pair production. The resulting electron is stopped directly in the detector and the positron annihilates almost immediately in one sector. In this process, two \SI{511}{\kilo\electronvolt} photons are emitted, which can escape or be detected in the detector. 
If both photons escape the detector, a so-called double escape peak (DEP) becomes visible. If, however, only one photon escapes the detector while the other is detected, a single escape peak (SEP) is observed. Those and further features are visible in the figure and are described below.

\begin{description}
    \item[FEP det:] The full energy of the \SI{2615}{\kilo\electronvolt} photon is deposited in the detector.
    \item[SEP det:] A \SI{2615}{\kilo\electronvolt} photon interacts via pair production in the detector and one photon escapes the detector.
    \item[DEP det:] A \SI{2615}{\kilo\electronvolt} photon interacts via pair production in the detector and both photons escape the detector.
    \item[DEP sec1:] A \SI{2615}{\kilo\electronvolt} photon interacts via pair production in sector one and both photons escape sector one.

    \item[P1:] A \SI{2615}{\kilo\electronvolt} photon interacts via pair production in another sector than sector one. One of the photons escapes the detector and one is detected in sector one.
    \item[P2:] A \SI{2615}{\kilo\electronvolt} photon interacts via pair production in sector one and both photons escape the sector, but one is detected in another sector.

    \item[P3:] A \SI{2615}{\kilo\electronvolt} photon interacts via pair production in another sector than sector one. One of the photons is detected in sector one and the other in another sector. 
    \item[P4:] A \SI{2615}{\kilo\electronvolt} photon interacts via pair production in sector one. One of the photons escapes the sector, but is detected in another sector.

    \item[L1:] A \SI{2615}{\kilo\electronvolt} photon interacts via pair production in another sector than sector one. One of the photons escape the detector and one is Compton-scattered in sector one and than escapes the detector. 
    \item[L2:] A \SI{2615}{\kilo\electronvolt} photon interacts via pair production in sector one. One of the photons is detected in another sector than sector one and one is Compton-scattered in sector one and than escapes the detector. 

    \item[B1:] A \SI{2615}{\kilo\electronvolt} photon is nearly \SI{180}{\degree} Compton-scattered in another sector than sector one and the remaining energy is deposited in sector one.
    \item[B2:] A \SI{2615}{\kilo\electronvolt} photon is nearly \SI{180}{\degree} Compton-scattered in sector one and the remaining energy is deposited in another sector.
\end{description}

The clearly visible structure of those features in large volume quad-grid detectors highlights the potential of those detectors and offers new analysis techniques, such as requirements on the deposited energy in several sectors to reduce background.

\subsection{Suppression of multi-sector events}
An advantage of the quad-grid structure of the detectors under investigation is the possibility to identify events in which energy is deposited in more than one sector, e.g. for Compton-scattered photons. Because of the expected localized energy deposition of $0\nu\beta\beta$ decays, a possible way to reduce the background induced by $\gamma$-radiation is to remove all events with a multiplicity of $N > 1$. 

To get an estimate of the possible background reduction, a multi-sector factor $M_f$ can be calculated. This factor is defined as the total number of events in an area of one FWHM around a FEP divided by the number of events with the additional condition that the multiplicity $N = 1$. Thus, $M_f$ is always greater or equal to one and is a measure of the probability that a single events deposits energy in more than one sector. 
It is expected that the multi-sector factors $M_f$ increase with higher energy due to the decreasing ratio between the cross sections for photoelectric absorption and Compton-scattering with increasing energies. Thus, the probability for multi-site events is higher for higher energies.

\begin{figure}[!ht]
    \centering
    \includegraphics[width=0.99\columnwidth]{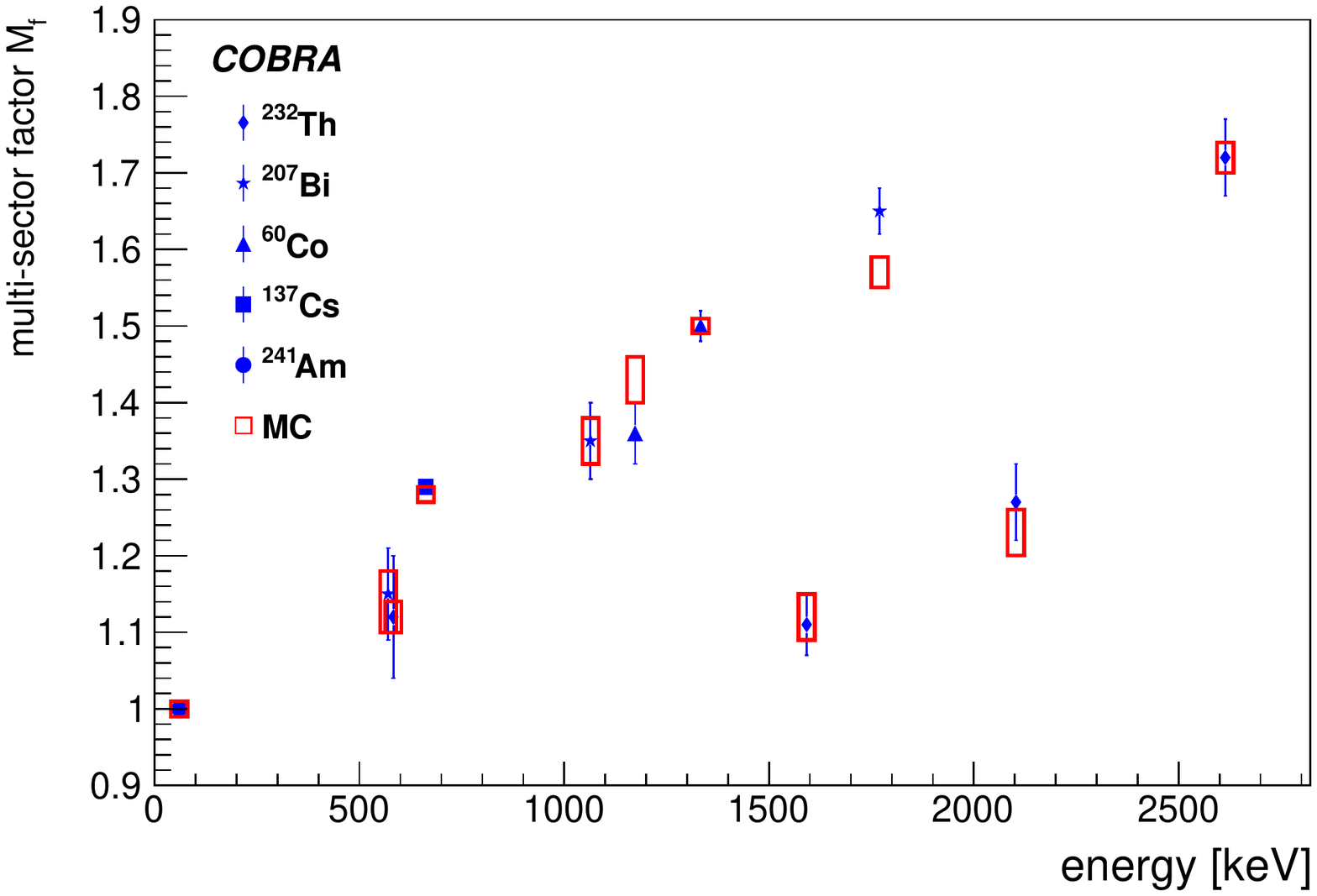}
    \caption{Multi-sector factors for FEPs from different sources. The measured values are shown in blue. The values from the simulations are shown in red. The different markers indicate different radioactive sources.}
    \label{fig:supression}
\end{figure}

$M_f$ was calculated for the measured data as well as for the simulated data. In Table \ref{tab:supression} and Figure \ref{fig:supression}, the multi-sector factor for several FEPs is shown. The uncertainties on $M_f$ arise from the fit to each peak as well as the numerical integration.
The results from the Monte Carlo simulations and the measurement are in good agreement within the uncertainties. 

\begin{table}[!ht]
    \centering
    \caption{Multi-sector factors $M_f$ for FEPs from different sources.}
    \begin{tabular}{rrlll}
        \toprule
        Source & E [keV]   & Note & $M_f (exp)$& $M_f (sim)$\\ 
        \midrule
        ${}^{241}$Am    & \num{60}  & -                 & \num{1.00\pm0.01}     & \num{1.00\pm0.01} \\ 
        ${}^{207}$Bi    & \num{570} & -                 & \num{1.15\pm0.06}     & \num{1.14\pm0.04} \\              
        ${}^{232}$Th    & \num{583} & -                 & \num{1.12\pm0.08}     & \num{1.12\pm0.02} \\ 
        ${}^{137}$Cs    & \num{662} & -                 & \num{1.29\pm0.01}     & \num{1.28\pm0.01} \\
        ${}^{207}$Bi    & \num{1064}    & -             & \num{1.35\pm0.05}     & \num{1.35\pm0.03} \\              
        ${}^{60}$Co     & \num{1173}    & -             & \num{1.36\pm0.04}     & \num{1.43\pm0.03} \\ 
        ${}^{60}$Co     & \num{1333}    & -             & \num{1.50\pm0.02}     & \num{1.50\pm0.01} \\          
        ${}^{232}$Th    & \num{1593}    & DEP           & \num{1.11\pm0.04}     & \num{1.12\pm0.03} \\
        ${}^{207}$Bi    & \num{1770}    & -             & \num{1.65\pm0.03}     & \num{1.57\pm0.02} \\               
        ${}^{232}$Th    & \num{2103}    & SEP           & \num{1.27\pm0.05}     & \num{1.23\pm0.03} \\ 
        ${}^{232}$Th    & \num{2615}    & -           & \num{1.72\pm0.05}     & \num{1.72\pm0.02} \\ 
        \bottomrule
    \end{tabular}

    \label{tab:supression}
\end{table}

The calculated factors of the FEPs rises from \numrange{1.0}{1.72} for energies from \SIrange{60}{2615}{\kilo\electronvolt}. 
The FEP at \SI{2615}{\kilo\electronvolt} is expected to produce many multi-site events and is thus suppressed in the $N = 1$ spectra. 
On the other hand, the DEP at \SI{1593}{\kilo\electronvolt} shows a smaller suppression than other lines of similar energy because of its localized energy deposition. Due to the underlying background from \isotope[228]{Ac} (\SI{1588}{\kilo\electronvolt}), \isotope[212]{Bi} (\SI{1621}{\kilo\electronvolt}) and the Compton continuum, the suppression is still larger then 1.0 even for the DEP. The SEP of the \SI{2615}{\kilo\electronvolt} $\gamma$-line does not fit to the trend of the multi-sector factors of the FEPs, either,  due to the underlying interaction process.

\section{Conclusion} \label{sec:conclusion_outlook}
A large \SI[product-units = power]{2x2x1.5}{\centi\meter} CPqG CdZnTe detector is operated and characterized.
It was shown that the energy reconstruction of small \SI{1}{\centi\meter\cubed} detectors is transferable to the larger detectors.
The measurements show a linear energy response with deviations smaller than 1\% for three of the sectors and for one sector it is smaller than 2\%. The detector has a good energy resolution and can be operated stably over the time period of at least one week. 
The CPqG design offers a simple and effective way to suppress multi-site events by using anti-sector coincidence and the multi-sector factors calculated with the data are in agreement with those predicted by the Monte Carlo simulations. 
The measurements indicate that the investigated detector meets requirements for usage in low-background operation and is the first step in building one layer of large CdZnTe detectors for COBRA.

\section*{Acknowledgments}
We thank the LNGS for the continuous support of the COBRA experiment. COBRA is supported by the Deutsche Forschungsgemeinschaft DFG.

\bibliography{cobra_characterization_6ccm_detectors}
\bibliographystyle{elsarticle-num}

\end{document}